\documentclass[aps,twocolumn]{revtex4}%
\usepackage{amssymb}
\usepackage{amsmath}
\usepackage{graphicx}%
\usepackage{amsfonts}
\usepackage{epsfig,psfrag,rotating,soul}
\providecommand{\U}[1]{\protect \rule{.1in}{.1in}}
\begin{document}
\title{Strong Gravity Approach to QCD and General Relativity}
\author{O. F. Akinto$\thanks{{}}$, Farida Tahir$\thanks{{}}$}
\affiliation{Department of Physics, COMSATS Institute of Information Technology, Islamabad, Pakistan}

\begin{abstract}
A systematic study of a Weyl type of action, which is scale free and quadratic
in the curvature, is undertaken. The dynamical breaking of this scale
invariance induces general relativity (GR) as an effective long distance limit
of the theory. We prove that the corresponding field equations of the theory
possess an effective pure Yang-Mills (i.e. QCD without quarks) potential,
which describes the asymptotic freedom and color confinement properties of
QCD. This inevitably leads to the solutions of \textbf{quantum Yang-Mills
existence on R}$^{4}$\textbf{\ (with its characteristic mass gap), and dark
matter problems}. The inherent \textbf{Bern-Carrasco-Johansson} (BCJ)
double-copy and gauge-gravity duality properties of this formulation lead to
the solutions of the \textbf{neutrino mass and dark energy problems}. This
approach provides a strong gravity basis for the unification of quantum
Yang-Mills theory (QYMT) with Einstein GR.


\  \  \  \  \  \  \  \  \  \  \  \  \  \  \  \  \  \  \  \  \  \  \  \  \  \  \  \  \  \  \  \  \  \  \  \  \  \  \  \  \  \  \  \  \  \  \  \  \  \  \  \  \  \ 

\end{abstract}
\maketitle

\section{Introduction}

\begin{quotation}
\footnotetext[1]{pheligenius@yahoo.com}\footnotetext[2]%
{farida\_tahir@comsats.edu.pk}\emph{"Who of us would not be glad to lift the
veil behind which the future lies hidden; to cast a glance at the next
advances of our science and at the secrets of its developments during future
centuries?" }$-$ \textbf{David Hilbert (1900).}

\emph{"It is by the solution of problems that the investigator tests the
temper of his steel; he finds new methods and new outlooks, and gains a wider
and freer horizon" }$-$ \textbf{David Hilbert (1900).}
\end{quotation}

In the early seventies Abdus Salam and his co-workers proposed the concept of
strong gravity, in which the successive self -interaction of a nonlinear
spin-2 field was used to describe a non-abelian field of strong interactions.
This idea was formulated in a two-tensor theory of strong and gravitational
interactions, where the strong tensor fields are governed by Einstein-type
field equations with a strong gravitational constant $G_{f}\approx10^{38}$
times the Newtonian constant $G_{N}$. Within the framework of this proposal,
tensor fields were identified to play a fundamental role in the
strong-interaction physics of quantum chromodynamics (QCD) \cite{CJI, ASJ,CSI,
DJS, YNE, ASCS}.

All the calculations done in the numerical lattice QCD and other related
experiments indicate that QCD, the worthy theory of strong interactions,
possesses gauge symmetry based on the group $SU(3)-$color of quantum
Yang-Mills theory (QYMT). Gravitational interactions also have similar
symmetry (the coordinate invariance in a space-time manifold), but resist
quantization. \emph{This prevents physicists from constructing a quantum
theory of gravity based on the gauge principle, and also inhibits the direct
unification of gravity with strong interaction} \cite{IANC}.

The origin of the difficulties is now clear to us: QCD action is scale
invariantly quadratic in the field strengths $F_{\mu \nu}^{i}$
(i.e.\emph{non-unitary}) and \emph{renormalizable,} while the Einstein-Hilbert
action for pure gravity is \emph{unitary} and \emph{nonrenormalizable}. Thus,
the unification of gravity with QCD seems unattainable; however, that is not
the case: The valiant attempt to disprove this \emph{prima facie}
impossibility offers an outstanding example of the inspiring effect which such
a very special and apparently important solution may have upon physics community.

\bigskip Having now recalled to mind the origin of the problem, let us turn to
the question of whether there is an existing unification scheme that can be
used to solve the problem. Strong gravity formulation is such the unification
scheme that allows the gravity to be merged with QYMT. In this case, a
gravitational action which possesses quadratic terms in the curvature tensor
has been shown to be renormalizable (\cite{KSS}, P.963 \& P.967). Here, the
resulting non-gauge-invariant divergences are absorbed by nonlinear
renormalizations of the gravitational fields and Becchi-Rouet-Stora
transformations (\cite{KSS}, P.953). In the following, \emph{the dynamical
breaking of the scale invariance of} \emph{Weyl action (which describes the
short distance behavior of strong gravity theory) induces: (1)
perturbative/short-range component of \ the\ non-relativistic QCD potential,
and non-relativistic quantum electrodynamic (QED) potential. (2) Einstein
general relativity as an effective long distance limit of the theory} $-$
\textbf{This is the \emph{fons et origo} of the gauge/gravity duality; and the
solution to the quantum Yang-Mills existence on R}$^{4}$ \textbf{and dark
matter problems, within the strong gravity formulation.}

The catch here is that quantum gravity (i.e. a quantum mechanically induced
gravity) cannot be derived straightforwardly by quantizing nonrenormalizable
Einstein GR but Weyl action which leads to Einstein's theory of gravity at
large distances\cite{IANC}; in the same way the gauge theory of
Glashow-Weinberg-Salam, $G_{EW}=SU(2)_{L}\times U(1)_{Y},$ reduces to
$U(1)_{Q}$ after the spontaneous symmetry breakdown\cite{EW, JCA}.

QCD possesses four remarkable properties that strong gravity must have for it
to be called a complete theory of strong interactions. The \  \textbf{first} is
\emph{asymptotic freedom} (i.e., the logarithmic decrease of the QCD coupling
constant $\alpha_{s}(Q_{0}^{2})\sim1/(\ln$ $Q_{0}^{2})$ at large momentum
transfers, or equivalently the decrease of $\alpha_{s}$ at small distances,
$\alpha_{s}(r)\sim1/(\ln r)$) which permits one to perform consistent
theoretical computations of hard processes using perturbation theory. This
property also implies an increase of the running coupling constant at small
momentum transfer, that is, at large distances. The \textbf{second} important
property is the\emph{\ confinement,} in which quarks and gluons are confined
within the domain of their strong interaction and hence cannot be observed as
real physical objects. The physical objects observed experimentally, at large
distances, are hadrons (mesons and baryons). The \textbf{third} characteristic
property is the \emph{dynamical breakdown of chiral symmetry}, wherein the
vector gauge theories with massless Dirac fermion fields $\psi$ are perfectly
chiral symmetric. However, this symmetry is broken dynamically when the vector
gauge theory is subjected to chiral $SU(2)$ rotations. This is the primary
reason why chiral symmetry is not realized in the spectrum of hadrons and
their low energy interactions\cite{BIF, QHN}. The \textbf{fourth} property is
the \emph{mass gap(}$\Delta$\emph{). Here, }every\emph{\ }excitation of the
QCD vacuum has minimum positive energy (i.e. $\Delta>0$); in other words,
there are no massless particles in the theory\cite{EW, JCA}. Additionally,

strong gravity must also be able to reproduce the two fundamental parameters
of QCD (i.e., coupling $\alpha_{s}$ and fundamental quark mass $m_{q}$
\cite{JBER}, P.178).

Thus, the three demands that must be met by strong gravity theory for it to be
called a unification scheme for QYMT-GR are:

\textbf{(1)} It must admit the four QCD properties afore-listed.

\textbf{(2)} It must be able recover the fundamental parameters of QCD (i.e.,
$\alpha_{s}$ and $m_{q}$).

\textbf{(3)} It must be able to reproduce Einstein's general relativity as the
limiting case of its long-distance behavior.

Any theory that fulfills these three demands can be termed "\textbf{a unified
theory of nature}"\textbf{.}

In the present paper, we study the structure of a dynamically broken
scale-invariant quantum theory (Weyl's action) within the context of strong
gravity formulation, and its general properties. The major problem which has
to be faced immediately is the unresolved question of unitarity of pure
gravity: Weyl's action is non-unitary while the Einstein-Hilbert action for
pure gravity is unitary. This problem is circumvented within the framework of
strong gravity: where the unitary Einstein-Hilbert term is induced after the
breakdown of the scale invariance of Weyl's action (\cite{ASCS}, P.324). To
put it in a proper and succinct context, Einstein GR emerges from the Weyl's
action after the dynamical breakdown of its scale invariance. Hence Einstein's
theory of gravity is not a fundamental theory of nature but the classical
output of the more fundamental \ gluon-dependent Weyl's action.

The paper is organized as follows.\textbf{\ In section II}, we briefly review
the BCJ double-copy construction of gravity scattering
amplitudes.\textbf{\ Section III} is devoted to the review of strong gravity
theory. Most importantly, we prove that BCJ double-copy construction exists
within the strong gravity formulation. The calculation of the dimensionless
strong coupling constant is done in the \textbf{section IV. }The\textbf{\ }%
theoretically obtained value is tested experimentally in the\textbf{\ section
V.} We present strong gravity as a massive spin-two theory in the
\textbf{section VI}. Here, we show that the dynamics of strong gravity theory
is fully symmetric, but its vacuum state is asymmetric. We also show in this
section that electroweak and custodial symmetries can be induced dynamically.
Critical temperature, fundamental mass and mass gap of the QCD vacuum are
obtained in the \textbf{section VII}. This leads to the derivation of the
effective pure Yang-Mills potential. The gauge-gravity duality property of
strong gravity theory is studied in the\textbf{\ section VIII}. We also show
that strong gravity possesses UV regularity and dynamical chiral symmetry
breaking in this same section. \ Confinement and asymptotic freedom properties
of the strong gravity is studied in the \textbf{section IX}. In this section,
we calculate the energy density of QCD vacuum. The existence of quantum
Yang-Mills theory on $R^{4}$ is established in the \textbf{section X}. The
vacuum stabilizing property of Higgs boson with mass $m_{H}=129GeV$ is studied
in \textbf{section XI.} The solutions to the neutrino mass, dark energy and
dark matter problems are presented in the \textbf{sections XII}, \textbf{XIII}
and \textbf{XIV} respectively. The physics of the repulsive gravity and cosmic
inflation is presented in the \textbf{section XV}. Conclusion is given in the
\textbf{section XVI.}

\section{Theoretical Preliminaries}

Research in strong gravity has always had a rather unique flavor, due to
conceptual difficulty of the field, and remoteness from experiment. We argue,
in this paper, that if the conceptual misconception $-$ namely, that gravity
is bedeviled with many untamable infinities $-$ that beclouds the field could
be circumvented, then the complexity enshrined in the field would become
highly trivialize.

The most powerful tool for removing this conceptual difficulty is encoded in a
long-known formalism: that the asymptotic states of gravity can be obtained as
tensor products of two gauge theory states (i.e. $gravity=gauge\otimes
gauge$). This idea was extended to certain interacting theories, in 1986, by
Kawai, Lewellen and Tye \cite{TOGO1}; and to strong-gravitational theory by A.
Salam and C. Sivaram in 1992 \cite{ASCS}. The modern understanding of this
double-copy formalism is largely due to the work of \textbf{B}ern,
\textbf{C}arrasco and \textbf{J}ohansson (BCJ). Formally, double-copy
construction (also known as \textbf{BCJ} construction) is used to construct a
gravitational scattering amplitude by using modern unitarity method, and the
scattering amplitudes of two gauge theory as building blocks \cite{TOGO2,
TOGO3}. This pathbreaking technique of computing perturbative scattering
amplitudes, which led to a deeper understanding of quantum field theory,
gravity, and to powerful new tools for calculating QCD processes, was awarded
the \emph{2014 J.J. Sakurai Prize for Theoretical Particle Physics}
\cite{TOGO4}.

\textbf{BCJ} construction has overturned the long-accepted dogma on Einstein's
GR, which posits that GR is nonrenormalizable. This new approach breaths new
life into the search for a fundamental unified theory of nature based on the
"supergravity" approach. Supergravity tries to tame the infinities encountered
in the Einstein's theory of gravity by adding "supersymmetries" to it. In a
variant of the theory called $N=8$ supergravity, which has eight new
"mirror-image" particles (gravitinos) allow physicists to tame the infinities
present in the Einstein's theory of gravity: other variants of supergravity
are $N=2,4$ Yang-Mills-Einstein-Supergravity (YMESG) and $N=0$
Yang-Mills-Einstein (YME) theories (\cite{TOGO5,TOGO6}, and the references
therein) $-$ Supergravity is like a \textbf{"young twig, which thrives and
bears fruit only when it is grafted carefully and in accordance with strict
horticultural rules upon the old stem"}.

As to the $N=0$ YME theory (where $N=0$ means that there are no
supersymmetries in the theory), we claim that this theory is by no means
different from the broken-scale-invariant Weyl's action. This assertion can
only be true if this action naturally possesses BCJ and guage-gravity duality
properties. The BCJ property is established in the next subsection, and we
show that the potential, carried by the broken-scale-invariant Weyl's action,
possesses this property in the \textbf{subsection D} of \textbf{section III}
of this paper. The gauge-gravity duality property of strong gravity is
established in \textbf{section VIII}: this is our \textbf{"guide post on the
mazy paths to the hidden truths"} of neutrino mass and dark energy problems.
The discovery made here is that both problems are connected by the effective
vacuum energy (or effective Weyl Lagrangian).

\subsection{Perturbative Quantum Gravity and Color/ Kinematics Duality: A
Review}

QCD (one of the variants of Yang-Mills theory) is the current well-established
theory of the strong interactions. Due to its asymptotic-free nature,
perturbation theory is usually applied at short distances; and the ensuing
predictions have achieved an astonishing success in explaining a wide range of
phenomena in the domain of large momentum transfers. Upon closer consideration
the question arises: Can perturbation theory be used to explore the quantum
behavior of gravity at short distances as well? The answer to that question is
a resounding yes! The discovery of \textbf{BCJ} principle is now our window
into the quantum world of gravity with tamable infinities at short distances.
This principle states that, regardless of the number of spacetime dimensions
and loops, a valid gravity scattering amplitude is obtained by replacing color
factors with kinematic numerators in a gauge-theory scattering amplitude. The
resulting gauge-coupling doubling is called \textbf{BCJ/double-copy} property
\cite{TOGO2, TOGO3}.

The gluon's scattering amplitudes, (in terms of cubic graphs) at L loops and
in D dimensions, are given by (\cite{TOGO2, TOGO3,TOGO5,TOGO6}, and the
references therein):%
\begin{equation}
A_{m}^{(L)}=i^{L-1}g_{\alpha}^{m-2+2L}\underset{i\text{ }\in \text{ }%
cubic}{\sum}\int \frac{d^{LD}\ell}{(2\pi)^{LD}}\frac{1}{S_{i}}\frac{c_{i}n_{i}%
}{D_{i}}%
\end{equation}

where $m$ is the number of points, $g_{\alpha}$ is the dimensionless gauge
coupling, $S_{i}$ are the standard symmetry factors and $D_{i}$ are
denominators encoding the structure of propagator in the cubic graphs. $c_{i}
$ are the color factors and $n_{i}$ are the kinematic numerators. BCJ
construction posits that within the gauge freedom of individual cubic graphs,
there exist unique amplitude representations that make kinematic factors
$n_{i}$ obey the same general algebraic identities as color factors. Hence,
color/kinematics duality holds: $n_{i}\iff c_{i}$ \cite{TOGO2, TOGO3}.

The double-copy principle then states that once the color/kinematics duality
is satisfied (i.e., $n_{i}\iff c_{i}$), the L-loop scattering amplitudes of a
supergravity theory (with $N\geq4$) are given by%
\begin{equation}
M_{m}^{(L)}=i^{L-1}\left(  \frac{k_{\alpha}}{2}\right)  ^{m-2+2L}%
\underset{i\text{ }\in \text{ }cubic}{\sum}\int \frac{d^{LD}\ell}{(2\pi)^{LD}%
}\frac{1}{S_{i}}\frac{n_{i}^{2}}{D_{i}}%
\end{equation}

where dimensionless $k_{\alpha}$ is the gravity coupling; and it is assumed
that the two involved gauge fields are from the same Yang-Mills theory. From
Eqs. (1) and (2), we have%
\begin{equation}
A_{m}^{(L)}=M_{m}^{(L)}\iff k_{\alpha}=2g_{\alpha}%
\end{equation}

Eq.(3), which is valid for all variants of supergravity with $N\geq4$, is the
expected gauge-coupling doubling or BCJ property. This property shows that
gravitons and gluons should be part of a fundamental unified theory of nature.

However, the devil is in the detail: the color-kinematics duality ($n_{i}\iff
c_{i}$) is more or less a conjecture; and the scattering-amplitude method of
probing the quantum nature of gravity is full of many mathematical
\emph{landmines}. Nevertheless, the conclusions of N = 8 supergravity theory
are indisputable. For we are convinced that the gauge-coupling doubling and
gauge-gravity duality should exist in the correct theory of quantum gravity
without appealing to supersymmetries. This is where strong gravity theory (or
point-like gravity) kicks in. Our present knowledge of the theory of strong
gravity puts us in a position to \emph{attack successfully }the problem of
\textbf{quantum gravity/point-like gravity} by using powerful-mathematical
tools (\emph{formula operators from differential geometry with their duality
and supersymmetry-like properties}) \emph{bequeathed} to us by
\emph{antiquity}.

We conclude this section with a great quote from one of the greatest
revolutionary mathematicians the world has ever known (\emph{David Hilbert})
\cite{TOGO7}: "If we do not succeed in solving a mathematical problem, the
reason frequently consists in our failure to recognize the more general
standpoint from which the problem before us appears only as a single link in a
chain of related problems. After finding this standpoint, not only is this
problem frequently more accessible to our investigation, but at the same time
we come into possession of a method which is applicable also to related
problems" $-$The \textbf{"standpoint"} discovered in this paper is the
\textbf{strong gravity theory}.

\section{Strong Gravity Theory: A Review}

We briefly review the standard formulation of strong gravity theory in this
section: (for more details see \cite{CJI, ASJ, CSI, DJS, YNE, ASCS, KSS} and
the references therein). Beginning with the two-gluon phenomenological fields
(i.e. double-copy construction), we re-establish strong gravity as a
renormalizable four-dimensional quantum gauge field theory by varying
\emph{Weyl action} with respect to the spacetime metric constructed out of the
two-gluon configuration. In this case, the two-point configuration (which
leads to the quantization of space-time itself) naturally introduces a minimum
length $2r_{g}$ (i.e. "intergluonic distance"); where $r_{g}$ is the "gluonic
radius". It should be emphasized here that this way of quantizing space-time
begins from the trajectories of \emph{two 2-gluons},i.e., curves or paths of
the geometry used. This method of constructing spacetime geometry from 2-gluon
phenomenology has been shown to be compatible with nature: The visualization
of the QCD vacuum (i.e.\textbf{visualization of action density of the
Euclidean-space QCD vacuum in three-dimensional slices of a }$24^{3}\times
36$\textbf{\ spacetime lattice}), by D. B. Leinweber, has shown that empty
space is not empty; rather it contains quantum fluctuations in the gluon field
at all scales (this is famously referred to as "gluon activity in a vacuum")
\cite{TOGO8}. This can only mean one thing: that gluon field is the
fundamental field of nature, and the spacetime metric/gravity is emergent from
2-gluon configuration. This is the main argument of BCJ/double-copy
construction. \emph{Simpliciter!}

By taking the vacuum states of hadron to be colorless (i.e. color-singlet),
the approximation of an external QCD potential (the hadron spectrum above
these levels) can be generated by color-singlet quanta. Based on the fully
relativistic QCD theory, these contributions have to come from the summations
of suitable Feynman diagrams in which dressed n-gluon configurations are
exchanged between several "flavors" of massless quarks. Thus, the simplest
such system (with contributions from n-gluon irreducible parts
$n=2,3,...,\infty$ and with the same Lorentz quantum numbers) will have the
quantum numbers of 2-gluon. The color singlet external field is then
constructed from QCD gluon field as a sum (\cite{DJS}, P.572):%
\begin{equation}
G_{\mu}^{a}G_{\nu}^{b}\eta_{ab}+G_{\mu}^{a}G_{\nu}^{b}G_{\sigma}^{c}%
d_{abc}+...
\end{equation}

where $\eta_{ab}$ is the $SU(3)_{C}$ color-metric, $d_{abc}$ is the totally
symmetric $8\otimes8\otimes8\rightarrow1$ coefficient and $G_{\mu}^{a}$ is the
dressed gluon field. The curvature would be generated by the derivatives of
$G_{\mu}^{a}$ (\cite{ASCS},P.323). The 2-gluon configuration can then be
written from Eq.(4) as%
\begin{equation}
g_{\mu \nu}(x)=G_{\mu}^{a}G_{\nu}^{b}\eta_{ab}%
\end{equation}

with
\begin{equation}
g=\det(g_{\mu \nu}(x))
\end{equation}

Eq.(5) is taken as the dominating configuration in the excitation systematics.
In this picture, the metric is constructed from a gluon-gluon interaction, and
the gluon-gluon effective gravity-like potential (effective Riemannian metric,
$g_{\mu \nu}$) would act as a metric field passively gauging the effective
diffeomorphisms (general coordinate transformations), just as is done by the
Einstein metric field for the general coordinate transformations of the
covariance group (\cite{YNE}, P.174).

It is crystal-clear that Eq.(5), as put forward by the proponents of strong
gravity, is by no means different from the double-copy structure of gauge
fields in the BCJ construction ($gravity=gauge\otimes gauge$); as such we
should be able to arrive at the same conclusions. The BCJ formalism
(double-copy construction) is formulated by using scattering-amplitude method.
Similarly, we show that double-copy construction can be obtained by using
formula operators from the differential geometry. Our approach puts BCJ
formalism on a proper mathematical footing: \emph{it puts flesh on the bones
of BCJ formalism.}

\subsection{Scale-Invariant-Confining Action for Strong Gravity Theory}

In analogy with the scale-invariant QCD action which is \emph{quadratic in the
field strengths} $F_{\mu \nu}^{i}$(with dimensionless coupling), we have the
corresponding Weyl action for gravity (\cite{ASCS}, P.322):%
\begin{equation}
I_{W}=-\alpha_{s}\int d^{4}x\sqrt{-g}C_{\alpha \beta \gamma \delta}C^{\alpha
\beta \gamma \delta}%
\end{equation}

where $\alpha_{s}$ is purely dimensionless and can be made into a running
coupling constant $\alpha_{s}(Q_{0}^{2}).$ It's worth noting that Eq.(7) is
not only \emph{generally covariant} but also \emph{locally scale invariant}
(\cite{IANC}, P.6). The Weyl's tensor ($C_{\alpha \beta \gamma \delta})$ is
constructed out of the corresponding Riemann curvature tensor, i.e., the
covariant derivatives involving gauge fields, characterized with the
generators of the conformal group. In the following, the metric is generated
by Eq.(5) (\cite{ASCS}, P.323).

The Weyl curvature tensor is defined as the traceless part of the Riemann
curvature \cite{TT}:%
\begin{align}
C_{\alpha \beta \gamma \delta}  & =R_{\alpha \beta \gamma \delta}-\frac{1}%
{n-2}(R_{\alpha \gamma}\eta_{\beta \delta}-R_{\alpha \delta}\eta_{\beta \gamma
}\nonumber \\
& -R_{\beta \gamma}\eta_{\alpha \delta}+R_{\beta \delta}\eta_{\alpha \gamma
})\nonumber \\
& +\frac{1}{(n-1)(n-2)}R(\eta_{\alpha \gamma}\eta_{\beta \delta}-\eta
_{\alpha \delta}\eta_{\beta \gamma})
\end{align}

Eq.(8) is constructed by using the trace-free property of Weyl tensor:%
\begin{equation}
\eta^{\alpha \gamma}C_{\alpha \beta \gamma \delta}=C_{\beta \alpha \delta}^{\alpha
}=0
\end{equation}

By contracting Eq.(8) with itself, we get%
\begin{align}
C_{\alpha \beta \gamma \delta}C^{\alpha \beta \gamma \delta}  & =R_{\alpha
\beta \gamma \delta}R^{\alpha \beta \gamma \delta}-\frac{4}{(n-2)}R_{\beta \delta
}R^{\beta \delta}\nonumber \\
& +\frac{2}{(n-2)(n-1)}R^{2}%
\end{align}

In \ four-dimension ($n=4$), Eq.(10) reduces to;%
\begin{equation}
C^{2}\equiv C_{\alpha \beta \gamma \delta}C^{\alpha \beta \gamma \delta}%
=R_{\alpha \beta \gamma \delta}R^{\alpha \beta \gamma \delta}-2R_{\beta \delta
}R^{\beta \delta}+\frac{1}{3}R^{2}%
\end{equation}

Thus, Eq.(7) becomes,%
\begin{equation}
I_{W}=-\alpha_{s}\int d^{4}x\sqrt{-g}(R_{\alpha \beta \gamma \delta}%
R^{\alpha \beta \gamma \delta}-2R_{\beta \delta}R^{\beta \delta}+\frac{1}{3}R^{2})
\end{equation}

\subsection{Gauss-Bonnet Invariant Theorem}

For space-time manifold topologically equivalent to flat space, the
Gauss-Bonnet theorem relates the various quadratic terms in the curvature as
\cite{IANC}:%
\begin{equation}
I_{GB}=-\alpha_{s}\int d^{4}x\sqrt{-g}(R_{\alpha \beta \gamma \delta}%
R^{\alpha \beta \gamma \delta}-4R_{\alpha \beta}R^{\alpha \beta}+R^{2})=0
\end{equation}

Using this property, we can rewrite Eq.(12) as%

\begin{equation}
I_{W}\longrightarrow I_{WGB}=I_{W}-I_{GB}=I_{W}%
\end{equation}

\begin{equation}
I_{W}=-2\alpha_{s}\int d^{4}x\sqrt{-g}\left[  R_{\beta \delta}R^{\beta \delta
}-\frac{1}{3}(R_{\gamma}^{\gamma})^{2}\right]
\end{equation}

where $R_{\beta \delta}$ is the Ricci tensor, which is a symmetric tensor due
to the Bianchi identities of the first kind, and its trace defines the scalar
curvature $R_{\gamma}^{\gamma}=R$ (\cite{SW-2}, P.153). By using Eqs.(7) and
(15), we have%
\begin{equation}
\int d^{4}x\sqrt{-g}\left(  R_{\beta \delta}R^{\beta \delta}-\frac{1}{3}%
R^{2}\right)  =\frac{1}{2}\int d^{4}x\sqrt{-g}C_{\alpha \beta \gamma \delta
}C^{\alpha \beta \gamma \delta}%
\end{equation}
Eq.(15) leads to the field equations \cite{TOGO9}:%

\begin{equation}
\sqrt{-g}g_{\mu \alpha}g_{\nu \beta}\frac{\delta I_{W}}{\delta g_{\alpha \beta}%
}=-\frac{1}{2}T_{\mu \nu}%
\end{equation}

Eq.(17) would be of fourth-order in the form (\cite{ASCS}, P.323):%

\begin{align}
& \frac{1}{2}g_{\mu \nu}(R_{\gamma}^{\gamma})_{;\delta}^{;\delta}+R_{\mu \nu}%
{}_{;\delta}^{;\delta}-R_{\mu;\nu;\delta}^{\delta}-R_{\nu;\mu;\delta}^{\delta
}-2R_{\mu \delta}R_{\nu}^{\delta}+\frac{1}{2}g_{\mu \nu}R_{\gamma \delta
}R^{\gamma \delta}\nonumber \\
& -\frac{1}{3}[2g_{\mu \nu}(R_{\gamma}^{\gamma})_{;\delta}^{;\delta
}-2(R_{\gamma}^{\gamma})_{;\mu;\nu}-2R_{\gamma}^{\gamma}R_{\mu \nu}+\frac{1}%
{2}g_{\mu \nu}(R_{\gamma}^{\gamma})^{2}]\nonumber \\
& =\frac{1}{4\alpha_{s}}T_{\mu \nu}%
\end{align}

The corresponding fourth-order Poisson equation and its linearized solution
are given as(\cite{ASCS}, P.323 \& 325):%
\begin{align}
\delta_{s}\nabla^{4}V  & =km_{0}\delta^{3}(r)\nonumber \\
V(r)  & =\alpha r
\end{align}

It is clear from Eq.(18) that its left-hand side vanishes whenever $R_{\mu \nu
}$ is zero (the vanishing of a tensor is an invariant statement (\cite{SW-2}%
,P.146)), so that any vacuum solution of Einstein equations would also satisfy
the ones from the quadratic action. A complete exact solution of the field
Eq.(18) (with metric signature $+---$) for a general spherical symmetric
vacuum metric is given as (\cite{ASCS}, P.323-324):%
\begin{equation}
ds^{2}=\alpha dt^{2}-\beta dr^{2}-r^{2}d\theta^{2}-r^{2}\sin^{2}\theta
d\phi^{2}%
\end{equation}

where
\begin{equation}
\alpha=1-\frac{\lambda_{1}}{r}-\lambda_{2}r-\lambda_{3}r^{2}%
\end{equation}

\begin{equation}
\beta=\left[  \alpha \right]  ^{-1}%
\end{equation}

$\lambda_{1},$ $\lambda_{2},$ and $\lambda_{3}$ in Eq.(21) are suitable
constants, related to the coupling constant. Dimensional analysis and natural
unit formalism then tell us that coupling constant ($\alpha)$ would remain
dimensionless provided that $\lambda_{1}$ carries the dimension of distance
([L]$GeV^{-1}$), $\lambda_{2}$ the dimension of mass ([M]$GeV$), and
$\lambda_{3}$ the dimension of squared mass ([M]$^{2},$ $GeV^{2}$). If we take
the mass to be the mass of the quark ($m_{q}$), then we can rewrite Eq.(21) as%

\begin{equation}
\alpha_{s}=1-\frac{\lambda_{1}}{r}-m_{q}r-m_{q}^{2}r^{2}%
\end{equation}

For the pure Yang-Mills theory (i.e. QCD without quarks), $m_{q}\rightarrow0 $
and Eq.(23) reduces to%
\begin{equation}
\alpha_{s}=1-\frac{\lambda_{1}}{r}%
\end{equation}

Based on the strong gravity theory and the formalism of the vacuum solution of
Einstein field equations \cite{CSI,SW-2,CWKJ}, $\lambda_{1}=G_{f}$ $m$.

With this value, Eq.(24) reduces to%
\begin{equation}
\alpha_{s}=g_{00}=1-\frac{G_{f}\text{ }m}{r}%
\end{equation}

and Eq.(20) becomes%
\begin{align}
ds^{2}  & =\left(  1-\frac{G_{f}\text{ }m}{r}\right)  dt^{2}-\left(
1-\frac{G_{f}\text{ }m}{r}\right)  ^{-1}dr^{2}\nonumber \\
& -r^{2}d\theta^{2}-r^{2}\sin^{2}\theta d\phi^{2}%
\end{align}

where mass $m$ is the only allowed mass in the theory, and is due to the
self-interaction of the two gluons (glueball). Eq.(26) is the well celebrated
Schwarzschild vacuum metric except that instead of normal Newtonian
gravitational constant ($G_{N}\approx10^{-19}GeV^{-1}$), we have
strong-gravitational constant ($G_{f}\approx1GeV^{-1}$).

\subsection{Broken Scale Invariance and Perturbative/Short Distance Behavior}

Once we have $\Lambda_{QCD}\equiv G_{f}^{-1}\approx1GeV$, the scale invariance
would be broken. An additional Einstein-Hilbert term linear in the curvature
would be induced, but the full action would still preserve its general
coordinate invariance (\cite{ASCS},P.324):%

\begin{equation}
I_{eff}=-\int d^{4}x\sqrt{-g}\left(  \alpha_{1}R_{\mu \nu}R^{\mu \nu}-\alpha
_{2}R^{2}+k^{-2}\alpha_{3}R\right)
\end{equation}

Here the induced Einstein-Hilbert term incorporates the phenomenological term
$1/k^{2}=\frac{1}{32\pi G_{N}}$ (\cite{KSS}, P.954 \& 967): this term is
called graviton propagator/ "pure Yang-Mills" propagator . By comparing
Eq.(27) with Eq.(15), we have%
\begin{align}
\alpha_{1}  & =\alpha_{3}=2\nonumber \\
\alpha_{2}  & =\frac{2}{3}%
\end{align}

Using natural units formalism, we can write
\begin{equation}
k^{-2}=\frac{1}{32\pi G_{N}}\approx1\times10^{17}GeV
\end{equation}

where $G_{N}\approx10^{-19}GeV^{-1}$ (in natural units) (\cite{ASJ}, P. 2668).

Eq.(27) gives rise to the mixture of fourth-order and second-order field
equations(\cite{ASCS}, P.324), whose solutions for the field of a
\textbf{localized mass} involves Yukawa and the normal $1/r$ potential terms.%
\begin{equation}
\alpha \nabla^{4}V+\beta \nabla^{2}V\approx km_{0}\delta^{3}(r)
\end{equation}

The corresponding solution of the Eq.(30) for a \textbf{point mass source} is
given as (\cite{VDS}, P. 3):%
\begin{equation}
V(r)=\frac{C_{1}}{r}-\frac{C_{2}}{r}e^{-\beta_{1}/r}+\frac{C_{3}}{r}%
e^{-\beta_{2}/r}%
\end{equation}

where $C_{1}=k^{2}M/8\pi \alpha_{3},$ $C_{2}=k^{2}M/6\pi \alpha_{3},$
$C_{3}=k^{2}M/42\pi \alpha_{3,\text{ }}\beta_{1}=\left[  \alpha_{3}%
^{1/2}(\alpha_{1}k^{2})^{-1/2}\right]  \times G_{f}^{3/2},$ and $\beta
_{2}=\alpha_{3}^{1/2}\left[  2\left(  3\alpha_{2}-\alpha_{1}\right)
k^{2}\right]  ^{-1/2}\times G_{f}^{3/2}.$ $M$ is unknown invariant mass (but
we identified it to be the invariant mass of the final hadronic state of the
theory, $M\equiv m$ (because final observable particle state must be color singlet)).

By using Eqs.(28) and (29), $\beta_{2}=\infty$ and thus Eq.(31) reduces to%
\begin{equation}
V(r)=\frac{C_{1}}{r}-\frac{C_{2}}{r}e^{-\beta_{1}/r}%
\end{equation}

\begin{align}
C_{1}  & =\frac{k^{2}m}{16\pi}\nonumber \\
C_{2}  & =\frac{k^{2}m}{12\pi}\nonumber \\
\beta_{1}  & =k^{-1}G_{f}^{3/2}%
\end{align}

As expected, the resulting infinity $\beta_{2}=\infty$ is tamed by the
nonlinear nature of the Weyl's action.

From Eqs.(32) and (33), we have
\begin{equation}
V(r)=\frac{k^{2}\text{ }m}{16\pi r}\left(  1-\frac{4}{3}e^{-\beta_{1}%
/r}\right)
\end{equation}

Eq.(32) is the exact equation obtained for the broken scale invariance and
perturbative behavior of strong gravity in\emph{\ }(\cite{ASCS}, P.325).

\subsection{Double-copy Construction in Strong Gravity}

From Eq.(34), we can write%
\begin{equation}
V(r)=\frac{k_{\alpha}^{2}\text{ }}{16\pi r}C
\end{equation}

where the dimensionless gravity coupling $k_{\alpha}^{2}\equiv k^{2}$ $m=32\pi
G_{N}\times m$ and $C\equiv1-\frac{4}{3}e^{-\beta_{1}/r}$ is the
"group-theoretic constant" of strong gravity theory.

It is to be recalled that the interaction energy, to the leading order, of two
static (i.e., symmetric) color sources of QCD without quarks (pure Yang-Mills
theory) is given by \cite{TOGO10,TOGO11,TOGO12,TOGO13}:%
\begin{equation}
E(r)=\frac{g_{\alpha}^{2}\text{ }}{4\pi r}C
\end{equation}
Where dimensionless gauge coupling $g_{\alpha}^{2}\equiv g^{2}(r)\times
m_{rg}$, and $m_{rg}$ is an arbitrary renormalization group scale formally
invoked, in quantum field theory, to keep the scale-dependent gauge coupling
($g^{2}(r)$) dimensionless. Since Eq.(35) is also the energy of two
interacting gluons, we can write (from Eqs.(35) and (36))%
\begin{equation}
V(r)=E(r)\iff k_{\alpha}=2g_{\alpha}%
\end{equation}

Eq.(37) is the required BCJ property. We have therefore proved the existence
of double-copy construction in strong gravity. It is remarkable to note that
despite different approaches taken by supergravity (scattering amplitude
method) and strong gravity (effective potential method), we still arrive at
the same conclusion (see Eqs. (3) and (37)).

\section{ QCD Evolution}

The body of experimental data describing the strong interaction between
nucleons (which is the non-perturbative aspect of QCD for $r\longrightarrow
\infty$) is consistent with a strong coupling constant behaving as $\alpha
_{s}\approx1$ \cite{TOGO14}: obviously this aspect of QCD is consistent with
the Eq.(25) for $r\longrightarrow \infty.$

One of the discoveries about \textbf{strong force} is that it diminishes
inside the nucleons, which leads to the free movement of gluons and quarks
within the hadrons. The implication for the strong coupling is that it drops
off at very small distances. This phenomenon is called "asymptotic freedom" or
\textbf{perturbative aspect of QCD}, because gluons and massless quarks
approach a state where they can move without resistance in the tiny volume of
the hadron \cite{TOGO15}. Hence for the strong gravity to describe the
perturbative aspect of QCD correctly, it must reproduce the value of strong
coupling constant $\alpha_{s}$ (\emph{by using the observed properties of
gluons: the mediators of strong force}) that is compatible with the
experimental data. This is what we set out to do in this section.

\subsection{Gluon Density}

The first thing to note here is that gluon, being a bosonic particle, obeys
Bose-Einstein statistics. The Fermi-Dirac and Bose-Einstein distribution
functions are given as (\cite{PCR}, P. 115);%
\begin{equation}
\aleph_{r}=\frac{g_{r}}{e^{\sigma_{1}+\sigma_{2}\in_{r}}\pm1}%
\end{equation}

where the positive sign applies to fermions and the negative to bosons.
$\aleph_{r}$ is the number of particles in the single-particle states, $g_{r}
$ is the degenerate parameter, $\sigma_{1}$ is the coefficient of expansion of
a gas of\textbf{\ weakly coupled particles} (\textbf{an ideal configuration
for describing the asymptotic freedom/perturbative regime of QCD}) inside the
volume $V$. $\sigma_{2}$ is the Lagrange undetermined multiplier and $\in_{r}$
is energy of the $r$-$th$ state. The value of $"\sigma_{1}"$ for boson gas at
a given temperature is determined by the normalization condition (\cite{PCR},
P. 112 and 115);%
\begin{equation}
N=\underset{r}{%
{\displaystyle \sum}
}\frac{g_{r}}{e^{\sigma_{1}+\sigma_{2}\in_{r}}-1}%
\end{equation}

The summation sign in Eq.(39) can be converted into an integral, because for a
particle in a box, the states of the system have been found to be very close.
Using the density of single-particle states function, Eq.(39) reduces to;%
\begin{equation}
N=\underset{0}{\overset{\infty}{\int}}\frac{D(\in)d\in}{e^{\sigma_{1}%
+\sigma_{2}\in}-1}%
\end{equation}

where $D(\in)d\in$ is the number of allowed states in the energy range $\in$
to $\in+d\in$ and $\in$ is the energy of the single-particle state. Using the
density of states as a function of energy, we have (\cite{PCR}, P. 290);%
\[
D(\in)d\in \text{ }=\frac{4\pi V}{h^{3}}2m\in \left(  \frac{m}{p}\right)  d\in
\]

with%
\[
p=\sqrt{2m\in}%
\]%
\begin{equation}
D(\in)d\in \text{ }=2\pi V\left(  \frac{2m}{h^{2}}\right)  ^{3/2}\in^{1/2}d\in
\end{equation}

where $p$ is the momentum of particle, $m$ its mass and $h$ is the Planck
constant. By putting Eq.(41) into Eq.(40), we have%
\begin{equation}
N=2\pi V\left(  \frac{2m}{h^{2}}\right)  ^{3/2}\underset{0}{\overset{\infty
}{\int}}\frac{\in^{1/2}d\in}{e^{\sigma_{1}+\sigma_{2}\in}-1}%
\end{equation}

but $\sigma_{1}=\sigma_{2}\times \mu_{eff}$ and $\sigma_{2}=1/kT.$ $\mu_{eff}$
is the effective potential, $k$ is the Boltzmann constant and $T$ denotes
temperature (\cite{PCR}, P.116). Since there is no restriction on the total
number of bosons (gluons), the effective potential is always equals to zero
($\mu_{eff}=0$) (this is true for the case where the minimum of the effective
potential continuously goes to zero as temperature grows\cite{CSB}). Thus,
Eq.(42) reduces to;%
\begin{equation}
N=2\pi V\left(  \frac{2m}{h^{2}}\right)  ^{3/2}\underset{0}{\overset{\infty
}{\int}}\frac{\in^{1/2}d\in}{e^{\in/kT}-1}%
\end{equation}

By using the standard integral(where $\varsigma(z)$ is the \textbf{Riemann
zeta function} and $\Gamma(z)$ is the \textbf{gamma function})%
\begin{equation}
\underset{0}{\overset{\infty}{\int}}\frac{x^{z-1}dx}{e^{x}-1}=\varsigma
(z)\Gamma(z)
\end{equation}
\bigskip

Eq.(43) becomes%
\begin{equation}
N=2.61V\left(  \frac{2\pi mkT}{h^{2}}\right)  ^{3/2}%
\end{equation}

Using $m=E/c^{2}$ and the average kinetic energy of boson gas in
three-dimensional space $E=3kT/2,$ Eq. (45) reduces to;%
\begin{equation}
\frac{N}{V}=\left[  \frac{(2.61)(3\pi)^{3/2}k^{3}}{(hc)^{3}}\right]  T^{3}%
\end{equation}

Define $n_{g}\equiv \frac{N}{V}$ and $\Xi \equiv \left[  \frac{(2.61)(3\pi
)^{3/2}k^{3}}{(hc)^{3}}\right]  =2.522\times10^{7}(mK)^{-3}.$ Hence the gluon
density ($n_{g}$) can be expressed as;%
\begin{equation}
n_{g}=\Xi T^{3}%
\end{equation}

Eq.(47) is the required result for the finite temperature and density relation
for gluon.

\subsection{Strong-gravity Coupling Constant}

The principle of general covariance tells us that the energy-momentum tensor
in the vacuum (with zero matter and radiation) must take the form;%
\begin{equation}
T_{00}=K\langle \rho \rangle
\end{equation}

Here $\langle \rho \rangle$ has the dimension of energy density and $K$
describes a real (strong-) gravitational field \cite{SER}. Hence Eq.(48)
reduces to;%
\begin{equation}
T_{00}=K(E_{vac})^{4}%
\end{equation}

and $K=g_{00}=C_{QCD}\times C_{grav}(strong-gravity$ coupling). $C_{QCD}$ is a
dimensionless coefficient which is entirely of QCD origin and is related to
the definition of QCD on a specific finite compact manifold. Similarly,
$C_{grav}$ is a dimensionless coefficient which is entirely of gravitational
origin \cite{SER,LZW,FRU,SWEN}. Therefore Eq.(49) becomes%
\begin{equation}
T_{00}=g_{00}(E_{vac})^{4}%
\end{equation}
Recall that energy density ($\rho_{vac}$) can also be written as%
\begin{equation}
\rho_{vac}=\frac{E_{vac}}{V}=V^{-1}\times E_{vac}%
\end{equation}

Eq.(51) is justified by the standard box-quantization procedure \cite{SER}.
Hence we have%
\begin{equation}
\rho_{vac}=n_{g}\times E_{vac}%
\end{equation}

where $n_{g}\equiv V^{-1}$ (number density)$.$

From the average kinetic energy for gas in three-dimensional space, we have
$T=2E_{vac}/3k.$ With this value, Eq.(47) reduces to%
\begin{equation}
n_{g}=\frac{8\Xi(E_{vac})^{3}}{27k^{3}}%
\end{equation}

Thus Eq.(52) becomes%
\begin{equation}
\rho_{vac}=\frac{8\Xi(E_{vac})^{4}}{27k^{3}}%
\end{equation}

Eq.(54) is the energy density of a single gluon. But based on double-copy
construction (see section II, Eqs.(3) and Eq.(37)), Eq.(54) is multiplied by
2, and thus,%
\begin{equation}
2\rho_{vac}=\frac{16\Xi(\Delta \varepsilon_{vac})^{4}}{27k^{3}}%
\end{equation}

Eq.(55) now represents two-point correlator-vacuum energy density. By
comparing Eq.(50) with Eq.(55), we have%
\[
\alpha_{s}=g_{00}=\frac{16\Xi}{27k^{3}}=2.336\times10^{19}(meV)^{-3}%
\]

As $1m=5.070\times10^{15}GeV^{-1}$, the above equation leads to%

\begin{equation}
\alpha_{s}=g_{00}=C_{QCD}\times C_{grav.}=0.1797
\end{equation}

Eq.(56) is the required strong (-gravity) coupling constant at the starting
point of QCD evolution. In the next section, we show the compatibility of
Eq.(56) with the perturbative QCD, which is the theory that describes
asymptotic freedom regime analytically.

\section{Perturbative Quantum Chromodynamics}

Computations in perturbative QCD are formally based on three conditions:
\textbf{(1)} that hadronic interactions become weak at small invariant
separation $r\ll \Lambda_{QCD}^{-1}$; \textbf{(2)} that the perturbative
expansion in $\alpha_{s}(Q_{0}^{2})$ is well-defined mathematically;
\textbf{(3)} factorization dictates that all effects of collinear
singularities, confinement, non-perturbative interactions, and the dynamics of
bound state can be separated constituently at large momentum transfer in terms
of (process independent) structure functions $G_{i/H}(x,Q),$ hadronization
functions $D_{H/i}(z,Q),$ or in the case of exclusive processes, distribution
amplitudes $\phi_{H}(x_{i},Q)$ \cite{AHM,GPL}. The asymptotic freedom property
of perturbative QCD($\beta_{0}=11-(2/3)n_{f}$) is given as (\cite{ASO}, P. 1):%
\begin{equation}
\alpha_{s}(Q_{0}^{2})=\frac{4\pi}{\beta_{0}\ln(\frac{Q_{0}^{2}}{\Lambda^{2}}%
)}<0.2\text{ \ for }Q_{0}^{2}>20GeV^{2}%
\end{equation}

In the framework of perturbative QCD, computations of observables are
expressed in terms of the renormalized coupling $\alpha_{s}(\mu_{R}^{2})$.
When one takes $\mu_{R}$ close to the scale of the momentum transfer $Q_{0}$
in a given process, then $\alpha_{s}(\mu_{R}^{2}\sim Q_{0}^{2})$ is indicative
of the effective strength of the strong interaction in that process. Eq.(57)
satisfies the following renormalization group equation (RGE) \cite{GDISS}:%
\begin{equation}
\mu_{R}^{2}\frac{d\alpha_{s}}{d\mu_{R}^{2}}=\beta(\alpha_{s})=-(b_{0}%
\alpha_{s}^{2}+b_{1}\alpha_{s}^{3}+b_{2}\alpha_{s}^{4}+O(\alpha_{s}^{5}))
\end{equation}

with%
\begin{equation}
b_{0}=(33-2n_{f})/12\pi
\end{equation}%
\begin{equation}
b_{1}=(153-19n_{f})/24\pi^{2}%
\end{equation}%
\begin{equation}
b_{2}=(2857-\frac{5033}{9}n_{f}+\frac{325}{27}n_{f}^{2})/128\pi^{3}%
\end{equation}

where Eqs.(59-61) are referred to as the 1-loop, 2-loop and 3-loop
beta-function coefficients respectively. The minus sign in Eq.(58) is the
origin of asymptotic freedom, i.e., the fact that the strong coupling becomes
weak for hard processes. Eq.(58) shows that \emph{RGE} is dependent on the
correct value of a purely dimensionless strong coupling constant ( $\alpha
_{s}$). Thus the precise calculation of its value (without appealing to the
choice of renormalization scheme and scale choice $Q_{0}^{2}$) would be the
holy grail of perturbative QCD.

\subsection{Experimental Test}

We begin by reviewing the systematic study of QCD coupling constant from deep
inelastic measurements in (\cite{VGKA} and the references therein), where many
experimental data were collected and analyzed at the next-to-leading order of
perturbative QCD (see Tables 2,3 and 6 of \cite{VGKA}) by using deep inelastic
scattering ($DIS$) structure functions $F_{2}(x,Q^{2}).$In these experimental
results, we are more interested in the $\alpha_{s}(90GeV^{2})=0.1797$ (in the
Table 6 of \cite{VGKA}) obtained when the number of points is 613. This is the
exact value we obtained theoretically in Eq.(56). Hence, we have not only
demonstrated that the perturbative expansion for hard scattering amplitudes
converges perturbatively at $\alpha_{s}=\alpha_{s}(90GeV^{2})=0.1797$ but also
able to prove that QCD is a strong-gravity-derived theory: \textbf{an
astonishing discovery!}\emph{\ }We have also validated the asymptotic freedom
property of perturbative QCD given in Eq.(57): namely, that the starting point
of QCD evolution is $Q_{0}^{2}=90GeV^{2}$ for $\alpha_{s}=0.1797<0.2.$

Having tested Eq.(56) experimentally, we therefore proceed to rewrite the
renormalization group equation (Eq.(58) ) as:%

\begin{equation}
\beta(\alpha_{s})=-\left[  b_{0}(0.1797)^{2}+b_{1}(0.1797)^{3}+b_{2}%
(0.1797)^{4}+...\right]
\end{equation}

Eq.(62) is an echo of \ "composition independence or universality property" of
the coupling $\alpha_{s}$ to all orders in the perturbative expansion for hard
scattering amplitudes.

\section{Strong Gravity as a Massive Spin-two Theory}

In the Einstein's GR, the \emph{Schwarzschild vacuum} is the solution to the
Einstein field equations that describes the gravitational field generated by a
\textbf{spherically symmetric mass }$m$, on the assumption that
the\textbf{\ electric charge}, and \textbf{orbital angular momentum (}%
$L$\textbf{)} of the \textbf{mass\ are all zero }\cite{CWKJ}.

It turns out that the Schwarzschild vacuum solution of the Einstein field
equations can be understood in terms of the Pauli-Fierz relativistic wave
equations for massive spin-2 particles which would mediate a short-range
tensor force (\cite{CSI}, P. 117). It follows that the two interacting gluon
fields ($G_{\mu}^{a}$ and $G_{\nu}^{b})$ are considered to be \textbf{dressed
gluon fields of the gravitational field}, i.e., the colors of the gluon fields
are covered or hidden within the spacetime base-manifold ($\eta_{ab}$) of the
color $-$ $SU(3)$ principal bundle (\cite{DJS}, P. 572), thereby making the
observable asymptotic states of gravity to be color-singlet/color-neutral.
Hence the resulting glueball (massive particle formed as a result of the
self-interaction of two gluons) of the theory (with spherically symmetric mass
$m$ and quantum numbers $J^{PC}=2^{-+}$) would still have the total angular
momentum of $2$. The validity of this statement is proved by using the
well-known Pauli-Fierz relativistic wave equations for massive particles of
spin-2(\cite{CSI}, P. 124):%
\begin{equation}
\square \phi_{\mu \nu}+m^{2}\phi_{\mu \nu}=0
\end{equation}

\begin{equation}
\partial_{\mu}\phi^{\mu \nu}=0\text{ (coordinate gauge condition)}%
\end{equation}

\begin{equation}
\phi_{\mu}^{\mu}=0\text{ (conformal gauge condition)}%
\end{equation}

\begin{equation}
\phi_{\mu \nu}=\phi_{\nu \mu}\text{ (symmetric condition)}%
\end{equation}

For the symmetric condition (Eq.(66)), the coordinate gauge condition given in
Eq.(64) eliminates four out of the ten components of the wave function
$\phi_{\mu \nu}$ of the Eq.(63); and the condition given in Eq.(65) eliminates
one more, leaving 5 degrees of freedom:%

\begin{equation}
2S+1=D=5\Longrightarrow S=2
\end{equation}

As a result of the Eq.(67), the following is true: \emph{strong gravity, as a
massive spin-2 theory, has five degrees of freedom (}$D=5$\emph{).}

Recall that the parity (P) and charge (C) quantum numbers can be expressed by%
\begin{equation}
P=(-1)^{J+1}%
\end{equation}

\begin{equation}
C=(-1)^{J}%
\end{equation}

and
\begin{equation}
J=L+S
\end{equation}

where $J$ is the total angular momentum, $L$ is the orbital angular momentum
and $S$ is the spin.

Thus, for the Schwarzschild vacuum solution (i.e., $L=0$), we have
\begin{equation}
J^{PC}=2^{-+}%
\end{equation}

\textbf{Requiring instead that }$\phi_{\mu \nu}\neq \phi_{\nu \mu}$
(\textbf{antisymmetric condition})\textbf{, we would have obtained
}$2S+1=D=1\Longrightarrow S=0$\textbf{\ and }$J^{PC}=0^{-+}$, \textbf{which is
a pseudoscalar state. An important consequence of this discovery is that the
underlying dynamics of the strong gravity theory is fully symmetric (i.e.
}$\phi_{\mu \nu}=\phi_{\nu \mu}\Longrightarrow S=2$\textbf{\ ) but its
ground/vacuum state is asymmetric (i.e. }$\phi_{\mu \nu}\neq \phi_{\nu \mu
}\Longrightarrow S=0;$ \textbf{meaning that the vacuum state must have massive
spin-zero particle(s) }$-$ \textbf{glueball/meson with mass }$m$\textbf{):
this is a formal description of spontaneous symmetry-breaking phenomenon.}

\subsection{Effective Lagrangian of a Massive Spin-2 Theory}

By using effective field theory (EFT) and the property of strong gravity (as a
massive spin-2 theory, $D=5$), the effective Lagrangian of the theory is
characterized by \cite{TOGO16}:%
\begin{equation}
L=\underset{i}{\sum}\frac{O_{i}}{M_{X}^{d_{i}-4}}%
\end{equation}

where $O_{i}$ are operators constructed from the \textbf{light fields (with
light mass)}, and information on any \textbf{heavy degrees of freedom ( with
heavy mass M}$_{X}$\textbf{)} is encoded in the coupling $\frac{1}%
{M_{X}^{d_{i}-4}}$. For $i=1$, we have%
\begin{equation}
L=\frac{O_{1}}{M_{X}^{d_{1}-4}}%
\end{equation}

Using $D=d_{1}=5$ means that the operator $O_{1}$ must carry the dimension of
squared energy ($O_{1}\sim E^{2}$) for the effective Lagrangian to carry the
dimension of energy:%
\begin{equation}
L=\frac{E^{2}}{M_{X}}%
\end{equation}

Eq.(74) is the effective Lagrangian of the strong gravity theory. The
invariant mass/energy operator $E^{2}=p_{\mu}p^{\mu}=m^{2}$ is called a flat
space/Poincar\.{e} invariant. This is characterized by an irreducible
representation of the Poincar\.{e} group (with spin $J$ ), and can be used to
describe a composite field (\cite{CSI}, P. 133-137) with five intrinsic
degrees of freedom (i.e. $D=d_{1}=5$). The importance of this statement will
be made manifest in the next subsection.

\subsection{Groups of Motions in Strong Gravity Admitting Custodial and
Electroweak Symmetries}

The fundamental theorem in the theory of strong gravity (as a massive spin-2
theory) contains two statements, namely:

(1) Strong gravity is a pseudo-gravity (\cite{YNE}, P.173).

(2) Strong gravity, as a massive spin-2 field theory, has five degrees of
freedom. The first statement means that the strong gravity must have a
fundamental group $SO(n_{1},n_{2})$. The group $SO(n_{1},n_{2})$ is the
special real pseudo-orthogonal group in $n_{1}+n_{2}$ dimensions. This group
has a non-compact group that is isomorphic to a generalized rotation group
(involving spherical (with positive curvature) and hyperbolic (with negative
curvature) rotations) in $R^{n_{1},n_{2}}$. Its maximal compact subgroup is
given as $SO(n_{1})\times SO(n_{2}).$The second statement forces us to write
$n_{1}+n_{2}=5$.

From the Eq.(5), the dressed gluon field $G_{\mu}^{a}$ can be separated into
asymptotic-flat connection ($N_{\mu}^{a}$), i.e. the \textbf{constant
curvature} (zero-mode) of the field and the normal gluon field ($A_{\mu}^{a}
$): $G_{\mu}^{a}=N_{\mu}^{a}+A_{\mu}^{a}$ (\cite{DJS}, P.572 \& \cite{YNE},
P.174). By using the de Sitter group formalism for the spacetime of constant
curvature, the non-compact groups (de Sitter groups) for strong gravity are
$SO(4,1)$ and $SO(3,2)$. The group $SO(4,1)$ is associated with the spacetime
manifold of constant positive curvature (denoted by $S(+)$), representing
spherical rotations, and $SO(3,2)$ is associated with the manifold of constant
negative curvature (denoted by $S(-)$), representing hyperbolic rotations. The
two spaces are embedded in the manifold with signature ($+--$). The maximal
compact subgroups for the two non-compact groups are (\cite{CSI}, P. 132):%
\begin{equation}
SO(4)\times SO(1)\approx SO(4)\approx SU(2)\times SU(2)
\end{equation}

\begin{equation}
SO(3)\times SO(2)\approx SU(2)\times U(1)
\end{equation}

Eqs.(75) and (76) can be used to label left-right and isospin-hypercharge
symmetries respectively:%
\begin{equation}
SU(2)_{L}\times SU(2)_{R}%
\end{equation}%
\begin{equation}
SU(2)_{L}\times U(1)_{Y}%
\end{equation}

Eq.(77) is called custodial symmetry of the Higgs sector. This symmetry is
spontaneously broken to the diagonal/vector subgroup after the Higgs doublet
acquires a nonzero vacuum expectation value (VEV): $SU(2)_{L}\times
SU(2)_{R}\longrightarrow SU(2)_{V}$ \cite{TOGO17}. Eq.(78) is the
\textbf{electroweak gauge symmetry} of the Standard Model (SM) of particle physics.

To break the \textbf{electroweak symmetry} at the \textbf{weak scale} and give
mass to quarks and leptons, Higgs doublets (that can sit in either $5_{H}$ or
$\overline{5}_{H})$ are needed. The extra 3 states are color triplet Higgs
scalars. The couplings of these color triplets violate lepton and baryon
number, and also allows the decay of nucleons through the exchange of a single
color triplet Higgs scalar. In order not to violently disagree with the
non-observation of nucleon (e.g. proton) decay, the mass of the single color
triplet must be greater than $\sim10^{11}GeV$ \cite{TOGO18}. It is to be
remarked here that this heavy mass would not disallow the violation of lepton
and baryon number: this is the key to unlocking the mystery of neutrino mass
problem. \emph{We shall return to this a little later}.

If the composite light field (with its five independent components) in the
\textbf{subsection A} of \textbf{section VI} is taken to be the Higgs field,
transforming in five-dimensional representation (i.e. $5_{H}$), then nature
would be permanently cured of its \textbf{vacuum catastrophe disease}. In this
case the invariant mass/energy operator of the light field would now be taken
to be the VEV of the Higgs doublets (i.e. $E\equiv \upsilon=246GeV$), and the
heavy mass of color triplet Higgs scalar would be encoded in the coupling
$1/M_{X}^{d_{1}-4}=1/M_{X}$. Here $M_{X}$ is the heavy mass characteristic of
the symmetry-breaking scale of the high-energy unified theory \cite{TOGO19}.
Once the high-energy unified theory that is compatible with nature is found,
the value of $M_{X}$ will show up automatically. This is where pure Yang-Mills
propagator kicks in.

\subsection{Type-A 331 Model}

One of the beyond-SM's of particle physics is the $SU(3)_{C}\times
SU(3)_{L}\times U(1)_{X}$ \ or $331$ model, in which the three fundamental
interactions (i.e. electromagnetic, weak and strong interactions) of nature
are unified at a particular energy scale $M_{U}$. This model is formulated by
extending the electroweak sector of the SM gauge symmetry. The unification of
the three interactions occurs at the energy scale $M_{U}\approx1\times
10^{17}GeV$ in the type-A variant of this model. In this variant of the model,
the 331 symmetry is broken to reproduce the SM electroweak sector at the
energy scale of $M_{X}=1.63\times10^{16}GeV$ \cite{TOGO20}. It is apparent
from the Eq.(29) that $k^{-2}=M_{U}=1\times10^{17}GeV-$ this is not surprising
because \textbf{electromagnetic}, \textbf{strong} and \textbf{weak} nuclear
interactions are all variants of Yang-Mills interaction $-$ Hence the type-A
331 model is compatible with the nature and $M_{X}$ (which is identified as
the mass of the single color triplet Higgs scalar) $=1.63\times10^{16}GeV$.
Thus Eq.(74) becomes%
\begin{equation}
L=\frac{E^{2}}{M_{X}}=\frac{\left(  246GeV\right)  ^{2}}{1.63\times10^{16}%
GeV}=3.7\times10^{-3}eV
\end{equation}

and the symmetry-breaking pattern is%
\begin{equation}
SU(3)_{L}\times U(1)_{X}\overset{M_{X}}{\longrightarrow}SU(2)_{L}\times
U(1)_{Y}\overset{E^{2}}{\longrightarrow}U(1)_{Q}%
\end{equation}

It is to be emphasized that the calculated value in the Eq.(79) is purely
based on the principle of naturalness: a composite field with five independent
components, which occurs naturally out of the strong gravity formulation, is
\textbf{identified} as the Higgs field $H$ transforming in five-dimensional
representations ($5_{H}$). \  \textbf{As we shall soon show, Eq.(79) connects
the solution of the dark energy problem to the neutrino mass problem.}

The chain of symmetry-breakings in the Eq.(80) has varying energy scales but
the Lagrangian $L$ of the whole system remains invariant: \emph{the physics of
vacuum seems to obey effective field theory rather than quantum field theory.}

\section{Some Consequences of Strong Gravity and their Physical
Interpretations}

This section is entirely devoted to the consequences of strong gravity. In
this case, we show the hitherto unknown connection between hadronic size,
physical lattice size and gluonic radius ($r_{g}$). From this, we calculate
the second-order phase transition/critical temperature $T_{c},$ and the
fundamental hadron mass of QCD.

\subsection{Calculation of the Gluonic Radius and Second-order Phase
Transition Temperature}

The configuration \ at $T>T_{c}$ for mass of the glueball for pure $SU(3)_{C}
$ is shown in the \textbf{Fig.1 }\cite{TOGO21}. Where $2r_{g}$ is the
intergluonic invariant separation.$\ S$ and $P$ represent scalar and
pseudoscalar glueball / gauge fields respectively. This figure is a perfect
representation of \ 2-gluon phenomenological field. It is interesting to note
that \textbf{Fig.1} has exactly the same structure with one-loop graviton
self-energy diagram (\cite{KSS}, P. 955). This is not a mere coincidence, it
only shows the compatibility of Eq.(5) with the tetrad formulation of GR, and
the existence of double-copy construction in all the variants of quantum
gravity theory. In what follows, we will heavily rely on the correctness of
the \textbf{Fig.1} as the valid geometry for strong gravity theory from the
point of view of 2-gluon phenomenology (double-copy construction).

We can therefore rewrite\ Eq.(25) for $T_{c}$ and gluonic radius $r_{g}$ as%

\begin{equation}
T_{c}=\frac{\left[  1-\alpha_{s}\right]  r_{g}}{G_{f}}%
\end{equation}

\emph{\ }By using Eq.(56), Eq.(81) becomes%
\begin{equation}
T_{c}=\frac{0.8203r_{g}}{G_{f}}%
\end{equation}
We now calculate the value of $r_{g}$ by using the value of the momentum
transfer, at which $\alpha_{s}$ converges perturbatively (i.e., $Q_{0}%
^{2}=90GeV^{2}$): see subsection A of section V.

Recall that the energy-wavelength relation is given as%
\begin{equation}
Q_{0}=\frac{hc}{\lambda}%
\end{equation}

Based on the geometry of \textbf{Fig.1}, we can write its associated
wavelength as;%
\begin{equation}
\lambda=2\pi r_{g}%
\end{equation}

Hence Eq.(83) reduces to;%
\[
Q_{0}=\frac{hc}{2\pi r_{g}}%
\]%
\begin{equation}
r_{g}=\frac{\hslash c}{Q_{0}}%
\end{equation}

But $Q_{0}^{2}=90GeV^{2}\Longrightarrow$ $Q_{0}=9.487GeV$ and $\hslash$
$=6.582\times10^{-16}eVs.$ Thus Eq.(85) reduces to,%
\begin{equation}
r_{g}=2.08\times10^{-17}m
\end{equation}

\begin{figure}[ht!]
\begin{center}
\includegraphics[ height=4cm, width=6cm]{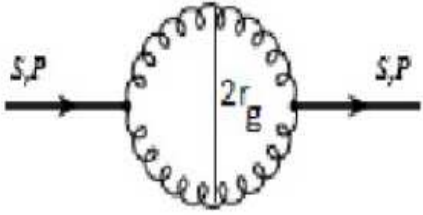}
\end{center}
\caption{Diagram for the contribution to the glueball (two-gluon) mass.}
\end{figure} 

Eq.(86) is the required gluonic radius. Clearly Eq.(86) is related to the
radius of hadron ($r_{h}$) \cite{CJI,ASCS,CSI,DJS,YNE}:%
\begin{equation}
r_{h}=10\times r_{g}%
\end{equation}

From the lattice QCD simulation performed at the initial run $\beta=2.2$ on a
$L^{3}T=24^{3}\times48$ lattice gives the physical lattice size ($L_{a}$) of
$2.08\times10^{-15}m$ \cite{AAKS}. By using Eq.(86), we can write%
\begin{equation}
L_{a}=10^{2}\times r_{g}%
\end{equation}

Hence Eqs.(86-88) show the connection between the gluonic radius, radius of
hadron and the physical lattice size.

It is generally believed that at sufficiently high temperature / density, the
QCD vacuum undergoes a phase transition into a chirally symmetric phase. Here,
the chirally symmetric phase transition will be second-order phase transition
$iff$ the conditions $T_{c}\neq0$ and $\mu_{eff}=0$ hold simultaneously
\cite{CSB}. Interestingly, we have \emph{a priori} claimed, during the
calculation of gluon density, that $\mu_{eff}=0$: an assertion that is
justified by the fact glueball, a self-conjugated particle with neutral color
and zero electric charge, has a vanishing effective/chemical potential (i.e.
$\mu_{eff}=0$) (\cite{TOGO19}, P.565). Thus the second-order chiral phase
transition temperature is calculated by using gluonic radius (Eq.(86)) , and
thus Eq.(82) becomes%
\begin{equation}
T_{c}=0.129GeV=129MeV
\end{equation}

where $G_{f}=10^{38}\times G_{N}=6.674\times10^{27}m^{3}kg^{-1}s^{-2}$ and
$1GeV=1.78\times10^{-27}kg.$

Hence, the chiral second-order phase transition in the strong gravity theory
occurs when $T_{c}=129MeV$ and $\mu_{eff}=0.$ Exactly the same values were
obtained in \cite{CSB, DSE} for second-order chiral phase transition in QCD
vacuum. We have thus established that strong gravity theory exhibits
second-order chiral symmetry in the limit of vanishing quark masses (
$m_{q}\rightarrow0$). It is worth noting here that the pure $SU(3)_{C}$ vacuum
metric (Eq.(26)), obtained in the limit $m_{q}\rightarrow0$, is compatible
with the glueball mass configuration given in the \textbf{Fig.1}, because
\textbf{Fig.1} was obtained in the limit of vanishing quark masses
\cite{TOGO21}.

\subsection{ Charmed Final Hadronic State of Strong Gravity}

Since we have shown that strong gravity theory possesses $SU(2)$ gauge field
(i.e. isospin symmetry, $SU(2)_{V}$) in the subsection B of section VI, it is
pertinent to investigate the structure of the fundamental mass formula of the theory.

In lattice QCD theory, the lattice spacing plays the role of ultraviolet
cutoff, since distances shorter than $"a"$ is not accessible. In the limit of
vanishing of quark masses ($m_{q}\rightarrow0$), this is the only dimensional
parameter and therefore all dimensionful quantities e.g. hadron and quark
masses will have to be given in units of the lattice spacing (\cite{QHN}, P.
271):%
\begin{equation}
m=\frac{1}{a}\text{ }f(\alpha(1/a),a)
\end{equation}

It is clear from Eq.(90) that the unknown function $f$ is dependent on the
strong coupling and lattice spacing. This equation is by no means different
from Eq.(25):%
\begin{equation}
m=\frac{r_{h}}{G_{f}}(1-g_{00})
\end{equation}

It is evident from Eqs.(90) and (91) that $\frac{1}{a}\equiv \frac{r_{h}}%
{G_{f}}$ and $f(\alpha(1/a),a)\equiv(1-\alpha_{s})=(1-g_{00}).$ By using
Eqs.(56) and (87), Eq.(91) becomes%
\begin{equation}
m=1.29GeV=1290MeV
\end{equation}

Eq.(92) is the fundamental, color-singlet mass scale of QCD vacuum.

The $\eta(1295)$ pseudoscalar state/ $\eta-meson$ state with $J^{PC}$
multiplets of $J^{PC}=0^{-+}$ has mass value of $m_{\eta}=1294\pm4MeV$
(\cite{TOGO18}, P.32). Similarly, the charm-quark (with charge $\frac{2}{3}$)
has the mass value ($m_{c}$) of $1.275\pm0.025GeV$ (\cite{TOGO18}, P.23). In
terms of the resumming threshold logarithms in the QCD form factor for the
B-meson decays to next-to-leading logarithmic accuracy, the mass formula for
the charm-quark is given as $m_{c}=m_{b}-m_{B}+m_{D}\approx1.29GeV$
\cite{UAG}. Where $m_{b},$ $m_{B}$ and $m_{D}$ denote bottom-quark, B- and
D-mesons respectively.

The correctness of the strong gravity theory in describing reality/nature is
clear from the above-quoted values. For we have shown in the \textbf{section
VI} of this paper that even though the underlying dynamics of the strong
gravity theory is fully symmetric ($\phi_{\mu \nu}=\phi_{\nu \mu}$), its vacuum
state is nonetheless asymmetric ($\phi_{\mu \nu}\neq \phi_{\nu \mu}$) with the
pseudoscalar quantum numbers $J^{PC}=0^{-+}$. In combining this fact with the
Eq.(92), the existence of the pseudoscalar $\eta$-$meson$ state $-$ with
$J^{PC}=0^{-+}$ and mass $m_{\eta}=1290MeV$ in the QCD vacuum $-$ is established.

\textbf{If} we take the dynamically induced coupling constant in the second
part of the Eq.(28) (i.e. $\alpha_{2}=\frac{2}{3}$) as the fundamental charge
of \ QCD vacuum $-$ attributed to the charm-quark $-$ (and taking into
consideration Eq.(92)), then we can say that charm-quark also exist in the QCD
vacuum. Thus, the fundamental quantities of the QCD vacuum are $\eta-meson$
(one of the examples of hadrons) and $charm-quark$. \emph{Based on this
understanding, we posit that the final hadronic state of strong gravity theory
is charmed (i.e. m = m}$_{\eta}=$\emph{\ m}$_{c}=1290MeV$\emph{).}

\textbf{In the next subsection, we establish the existence of mass gap within
the formulation of strong gravity (by using the vector sugroup (i.e. isospin
symmetry }$SU(2)_{V}$\textbf{) of the custodial symmetry in the Eq.(77)); and
also justify the validity of using the dynamically induced coupling constant
(}$\alpha_{2}=\frac{2}{3}$\textbf{) as the fundamental charge of the QCD
vacuum.}

\subsection{Mass Gap}

QCD is widely accepted as a dynamical quantum gauge theory of strong
interactions not only at the fundamental quark-gluon level, but also at the
hadronic level. In this picture, any color-singlet mass scale parameter must
be expressed in terms of the mass gap \cite{MGA}:%

\begin{align}
m  & =const\times m_{gap}\nonumber \\
m  & =const\times m_{gap}=1290MeV
\end{align}

where \emph{const.} denotes arbitrary constant.

In particle physics, particles that are affected equally by the strong force
but having different charges, such as protons and neutrons, are treated as
being different states of the same nucleon-particle with isospin values
related to the number of charge states:%
\begin{equation}
N=\left(
\begin{array}
[c]{c}%
N^{+}\\
N^{0}%
\end{array}
\right)  =\left(
\begin{array}
[c]{c}%
p\\
n
\end{array}
\right)
\end{equation}
The isospin symmetry ($SU(2)_{V}$) then demands that both charge states should
have the same energy in order to preserve the invariance of the Hamiltonian
(\textbf{H}) of the system. This means that isospin symmetry is a statement of
the invariance of \textbf{H }of the strong interactions under the action of
the Lie group $SU(2)$. However, the near mass-degeneracy of the neutron and
proton points to an approximate symmetry of the Hamiltonian describing the
strong interactions \cite{DGR, CITZ}. The mass gap ($m_{gap}$) $-$ which is
responsible for the approximate symmetry of strong interaction $-$ in this
case must be the energy difference between the proton state and neutron state
of the proton-neutron $SU(2)$ doublet fundamental representation (with gauged
isospin symmetry): $m_{gap}\equiv m_{n}-m_{p}\approx1.29MeV$. Where $m_{p}$
and $m_{n}$ are the masses of proton and neutron respectively (\cite{TOGO19}%
,P.152). It is to be noted here that $m_{n}-m_{p}$ is the transition
(excitation) energy needed to transform neutron into proton (\cite{SW-2},
P.548). In this picture, the mass gap is nothing but the energy difference
between these two states in the isospin space. From the foregoing, the
approximate $SU(2)_{V}$ isospin symmetry of the strong nuclear force is
dependent on the non-vanishing of $m_{gap}$, and hence the color-singlet mass
spectrum of the QCD matter must depend on it.

Thus Eq.(93) becomes%
\begin{equation}
m=10^{3}\times(m_{n}-m_{p})=1290MeV
\end{equation}

and
\begin{equation}
m_{gap}=m_{n}-m_{p}\approx1.29MeV
\end{equation}

It is to be recalled that the fundamental charge (of $U(1)$ and $SU(2)$ gauge
fields) is related to the electroweak coupling constants via the
Weinberg-Salam geometric relations: $e=g_{1}\cos \theta_{w}=g_{2}\sin \theta
_{w}$ and $\cos \theta_{w}=m_{W}/m_{Z}$ \cite{SW-3}. Where $g_{1}$ and $g_{2}$
are the gauge couplings of $U(1)$ and $SU(2)$ gauge fields respectively.
$\theta_{w}$ is the mixing angle, $e$ is the fundamental charge, $m_{W}$ is
the mass of $W$-boson and $m_{Z}$ is the mass of $Z$-boson. By using
$m_{W}=80.385GeV,m_{Z}=91.1876GeV$ \cite{SW-4}$,e=\alpha_{2}=2/3$, we have
$\theta_{w}\approx28.17^{0}$ and $g_{2}%
=0.6666666667/0.4720892507=1.4121623522.$ This is the nucleon coupling
constant for the two-flavor (i.e. proton and neutron) $SU(2)$ representation.
The value of $g_{2}$ ($=1.4121623522$) is to be compared with the nucleon
axial coupling constant computed from two-flavor $SU(2)$ lattice QCD:
$g_{A}=1.412(18)$ \cite{SW-5}.

In the next subsection, we demonstrate that the values of $m_{gap}$ and
$T_{c}$ do not only play a very important role in the Big Bang nucleosynthesis
but are also part of the primordial constituents of the QCD vacuum.

\subsection{Big Bang Nucleosynthesis (BBN)}

BBN refers to the production of relatively heavy nuclei from the lightest
pre-existing nuclei (i.e., neutrons and protons with $m_{gap}=1.29MeV$) during
the early stages of the Universe. Cosmologists believe that the necessary and
sufficient condition for nucleosynthesis to have occurred during the early
stages of the universe is that the value of equilibrium neutron fraction
($X_{n}$) or the neutron abundance must be close to the optimum value, i.e.,
$X_{n}\approx50\%$ (\cite{SW-2}, P.550). In fact, the value of $X_{n}$ at the
time $t=0$ was calculated to be $X_{n}=0.496=49.6\%$ (\cite{SW-2}, P.549).

The equilibrium neutron fraction for temperature $T\gtrsim3\times10^{10}K$ is
given as (\cite{SW-2}, P.550):%
\begin{equation}
X_{n}\approx \left[  1+e^{E/kT}\right]  ^{-1}%
\end{equation}

where $E=m_{gap}=1.29MeV.$ By using natural unit approach (i.e., setting the
Boltzmann constant $k=1$) and using the value of critical temperature
($T=T_{c}=129MeV$), Eq.(97) reduces to%
\begin{equation}
X_{n}\approx \left[  1+e^{0.01}\right]  ^{-1}=49.75\%
\end{equation}

The value in the Eq.(98) is compatible with the value obtained at the time
$t=0$ (i.e., $X_{n}=49.6\%$) , and is approximately equal to the optimum value
($X_{n}\approx50\%$). This can only mean two things: (i) $m_{gap}$ and $T_{c}$
existed at time $t=0$ of BBN processes. (ii) These two quantities are the
fundamental quantities of QCD / quantum vacuum.

According to the detailed calculations of Peebles and Weinberg, the abundance
by weight of cosmologically produced helium is given as (\cite{SW-2}, P.554):%
\begin{equation}
X_{H^{4}e}=2X_{n}%
\end{equation}

By combining Eqs.(98) and (99), we have%
\begin{equation}
X_{H^{4}e}=99.5\%
\end{equation}

Eq.(100) confirms the validity of Eq.(99), namely, that the total amount of
neutrons before nucleosynthesis must be equal to total amount of helium
abundance after the nucleosynthesis.

The threshold for the reaction $p+\overline{\nu}_{e}\rightarrow n+e^{+}$ is at
$m_{e}+m_{gap}=1.8MeV$ (\cite{SW-2}, P.544). Thus the mass of electron
($m_{e}$) is $m_{e}=0.51MeV.$

The invariance of the mass gap is supported by the following transitions
(\cite{SW-2}, P.548):%
\begin{align}
E_{e}-E_{\nu}  & =m_{gap}\text{ for }n+\nu \longleftrightarrow p+e^{-}%
\nonumber \\
E_{\nu}-E_{e}  & =m_{gap}\text{ for }n+e^{+}\longleftrightarrow p+\overline
{\nu}\nonumber \\
E_{\nu}+E_{e}  & =m_{gap}\text{ for }n\longleftrightarrow p+e^{-}%
+\overline{\nu}%
\end{align}

Eq.(101) clearly shows that mass gap is invariant under crossing-symmetry.

By using the values of $\alpha_{s}$ and $m$, we proceed to solve Eqs.(19) and
(34) completely. From Eq.(34), we have%
\begin{align}
F  & \equiv \frac{k^{2}\text{ }m}{16\pi}=\frac{32\pi G_{N}\times m}{16\pi
}\nonumber \\
F  & =2G_{N}\text{ }m=2.580\times10^{-19}%
\end{align}

Eq.(102) is to be compared with the ratio of the proton mass to the Planck
mass scale ($\frac{M_{proton}}{M_{Planck}}\approx10^{-19}$).

By using Eq.(29) and the value of $G_{f}$ ($\approx1GeV^{-1}$ \cite{ASJ}, P.
2668), the last part of Eq.(33) becomes%
\begin{equation}
\beta_{1}=3.162\times10^{8}GeV^{-1}%
\end{equation}

One of the properties of the confining force is the notion of "dimensional
reduction" which suggests that the calculation of a large planar Wilson loop
in $D=4$ dimensions reduces to the corresponding calculation in $D=2$
dimensions. In this case, the leading term for the string tension is derived
from the two-dimensional strong-coupling expansion (\cite{JEFF}, P.49-50).

Following this line of reasoning, $\alpha_{s}$ is made into a dimensionful
coupling (dimensional transmutation) as follows:%
\begin{align}
\sigma & \equiv \alpha_{s}[m]^{4-D}=0.1797\times(1.29GeV)^{2}\nonumber \\
\sigma & =0.299GeV^{2}%
\end{align}

Note that $\alpha_{s}$ is dimensionless (as expected) only in four dimensions,
but here we use $D=2$ in order to obtain the \textbf{Wilson-like}
\textbf{string tension} (which represents the geometry of the Weyl's action
because it is rotationally symmetric). Eq.(104), which is called string
tension, is to be compared with the value $\sigma=0.27GeV^{2}$ \cite{ANIVA}.
With these values, the confinning potential ($V_{conf}$)/linearly rising
potential in the Eq.(19) reduces to%

\begin{equation}
V_{conf}\text{ }(r)=\sigma r
\end{equation}

\bigskip and the perturbative aspect ($V_{pert}$) of strong gravity (Eq.(34)) becomes%

\begin{equation}
V_{pert}(r)=\frac{F}{r}-\frac{4}{3}\frac{(Fe^{-\beta_{1}/r})}{r}%
\end{equation}

Where the color factor ($C_{F}$)/Casimir invariant associated with gluon
emission from a fundamental quark $-$ present in the Eq.(106) $-$ for $SU(3)$
gauge group (with $N=3$) is given as
\begin{equation}
C_{F}=\frac{1}{2}\left(  N-\frac{1}{N}\right)  =\frac{4}{3}%
\end{equation}

and
\begin{equation}
e^{-\beta_{1}/r}=\underset{n=0}{\overset{\infty}{\sum}}\left(  -1\right)
^{n}\frac{\left(  \frac{\beta_{1}}{r}\right)  ^{n}}{n!}%
\end{equation}

Hence the effective pure Yang-Mills potential ($V_{YM}^{eff}$ $(r)$) of strong
gravity theory (from Eqs.(105) and (106)) is%
\begin{align}
V_{YM}^{eff}(r)  & =V_{pert}(r)+V_{conf}\text{ }(r)\nonumber \\
V_{YM}^{eff}(r)  & =\frac{F}{r}-\frac{4}{3}\frac{(Fe^{-\beta_{1}/r})}%
{r}+\sigma r
\end{align}

\section{GAUGE-GRAVITY DUALITY}

In this section, we show that strong gravity theory possesses gauge-gravity
duality property.

\subsection{NRQED and NRQCD Potentials}

The perturbative non-relativistic quantum electrodynamics (NRQED) that gives
rise to a \textbf{repulsive Coulomb potential between an electron-electron
pair} is due to one photon exchange, and this repulsive Coulomb potential is
given by \cite{TOGO22}:%
\begin{equation}
V_{QED}(r)=\frac{\alpha_{e}}{r}%
\end{equation}

where the QED running coupling $\alpha_{e}=\frac{\alpha(0)}{1-%
{\textstyle \prod}
(Q^{2})}$ . $\alpha(0)\approx1/137$ and $%
{\textstyle \prod}
(Q^{2})$ are the vacuum polarization insertions \cite{TOGO23}. Similarly, the
perturbative component of the NRQCD potential between two gluons or between a
quark and antiquark is given as \cite{TOGO22}:%
\begin{equation}
V_{QCD}(r)=-\frac{4}{3}\frac{\alpha_{s}(r)}{r}%
\end{equation}

where the strong running coupling $\alpha_{s}(r)$ must exponentiate in order
to account for the nonlinearity of the gluon self-interactions.

The total color-singlet NRQCD potential is (\cite{TOGO22}, P.273 \&
\cite{TOGO24}, P.39):
\begin{equation}
V_{QCD}(r)=-\frac{4}{3}\frac{\alpha_{s}(r)}{r}+kr
\end{equation}

Obviously, Eq.(106) contains both the NRQED potential (Eq.(110)) and NRQCD
potential (Eq.(111)). Hence the perturbative/short-range aspect of the strong
gravity theory (derived completely entirely from the broken-scale-invariant
Weyl's action in the Eq.(27)) unifies NRQED and NRQCD with one single coupling
constant $F$:%
\begin{equation}
V_{pert}(r)=F\left(  \frac{1}{r}-\frac{4}{3}\frac{e^{-\beta_{1}/r}}{r}\right)
\end{equation}

It is important to note that the QCD part (second term) of the Eq.(113) is
QED-like (first term) apart from the color factor 4/3 $-$ which shows that
there is more than one gluon $-$ and the exponential function $-$ which
accounts for the self-interaction between the gluons ( the \emph{fons et
origo} of nonlinearity in the Yang-Mills theory). Thus, strong gravity theory
is a \textbf{gauge theory}: we mention in passing that Eq.(112) is also
obtainable from the Eq.(109).

In the next subsection, we prove that the Einstein's theory of \textbf{gravity
}can also be derived from the same equation (Eq.(27)) that gave rise to the Eq.(106).

\subsection{\bigskip Effective Einstein General Relativity}

So far, we have been dealing with the short-range behavior of the strong
gravity theory. In this subsection, we take a giant step towards deriving the
Einstein GR entirely from the strong gravity formulation. To set the stage, we
rewrite Eq.(27) as:%
\begin{align*}
I_{eff}  & =-\int d^{4}x\sqrt{-g}\left(  \alpha_{1}R_{\mu \nu}R^{\mu \nu}%
-\alpha_{2}R^{2}\right)  -\\
& \int d^{4}x\sqrt{-g}k^{-2}\alpha_{3}R
\end{align*}

By using Eq.(28), the above equation becomes%
\begin{align}
I_{eff}  & =-\int d^{4}x\sqrt{-g}\left(  2R_{\mu \nu}R^{\mu \nu}-\frac{2}%
{3}R^{2}\right)  -\nonumber \\
& 2\int d^{4}x\sqrt{-g}k^{-2}R
\end{align}

\subsubsection{The Matter Action}

Without using any rigorous mathematics, we would like to show that the part of
the Eq.(114) containing the quadratic terms is in fact the matter action
($I_{M}$). From Eq.(16), we have%

\begin{align}
I_{M}  & \equiv \int d^{4}x\sqrt{-g}\left(  2R_{\mu \nu}R^{\mu \nu}-\frac{2}%
{3}R^{2}\right)  =\nonumber \\
& \int d^{4}x\sqrt{-g}C_{\mu \nu \alpha \beta}C^{\mu \nu \alpha \beta}\nonumber \\
I_{M}  & =\int d^{4}x\sqrt{-g}C_{\mu \nu \alpha \beta}C^{\mu \nu \alpha \beta}%
\end{align}

The fact that Weyl Lagrangian density ($C_{\mu \nu \alpha \beta}C^{\mu \nu
\alpha \beta}$) is a conserved quantity due to its general covariance property
means that we can write
\begin{equation}
\delta \left(  C_{\mu \nu \alpha \beta}C^{\mu \nu \alpha \beta}\right)  =0
\end{equation}

This ensures the conservation of energy-momentum.

By using the principle of stationary action on the Eq.(115) and taking
Eq.(116) into consideration, we have%
\begin{equation}
\delta I_{M}=\frac{1}{2}\int d^{4}x\sqrt{-g}\left(  g^{\mu \nu}C_{\mu \nu
\alpha \beta}C^{\mu \nu \alpha \beta}\right)  \delta g_{\mu \nu}%
\end{equation}

Recall that the energy-momentum tensor is defined as \cite{TOGO25}:%
\begin{equation}
T_{\mu \nu}\equiv \frac{-2\delta \left(  \sqrt{-g}%
\mathcal{L}%
_{M}\right)  }{\sqrt{-g}}=\frac{-2\delta%
\mathcal{L}%
_{M}}{\delta g^{\mu \nu}}+g_{\mu \nu}%
\mathcal{L}%
_{M}%
\end{equation}

where $%
\mathcal{L}%
_{M}$ is the matter conserved Lagrangian density. Using the Weyl conserved
Lagrangian density, we have%
\begin{equation}
T_{\mu \nu}=\frac{-2\delta \left(  C_{\mu \nu \alpha \beta}C^{\mu \nu \alpha \beta
}\right)  }{\delta g^{\mu \nu}}+g_{\mu \nu}\left(  C_{\mu \nu \alpha \beta}%
C^{\mu \nu \alpha \beta}\right)
\end{equation}

By using Eq.(116), Eq.(119) reduces to%
\begin{align}
T_{\mu \nu}  & =g_{\mu \nu}\left(  C_{\mu \nu \alpha \beta}C^{\mu \nu \alpha \beta
}\right)  \text{or}\nonumber \\
T^{\mu \nu}  & =g^{\mu \nu}\left(  C_{\mu \nu \alpha \beta}C^{\mu \nu \alpha \beta
}\right)
\end{align}

Clearly Eq.(120) is a conserved (due to Eq.(116)) symmetric (due to the
presence of $g^{\mu \nu}$) tensor (\cite{SW-2}, P. 360); and its nonlinearity
represents the effect of gravitation on itself. To deal with this nonlinear
effect, Principle of Equivalence is normally invoked, in which any point $X$
in an arbitrarily strong gravitational field is the same as a locally inertial
coordinate system such that $g_{\alpha \beta}(X)=\eta_{\alpha \beta}$
(\cite{SW-2}, P. 151).

Hence Eq.(117) becomes%
\begin{equation}
\delta I_{M}=\frac{1}{2}\int d^{4}x\sqrt{-g}T^{\mu \nu}\delta g_{\mu \nu}%
\end{equation}

Eq.(121) is the equation of energy-momentum tensor for a material system
described by matter action \cite{SW-2}.

\subsubsection{Pure Gravitational Action}

By using the value of \ $k^{-2}(=\frac{1}{32\pi G_{N}})$ from Eq.(29), the
linear term part of the Eq.(114) is written as%
\begin{align}
I_{G}  & =-2\int d^{4}x\sqrt{-g}k^{-2}R\nonumber \\
I_{G}  & =-\frac{1}{16\pi G_{N}}\int d^{4}x\sqrt{-g}R
\end{align}

We can therefore write%

\begin{equation}
I_{eff}=I_{M}+I_{G}%
\end{equation}

By using the general covariance property of Weyl's action, we can write
\begin{equation}
\delta I_{eff}=\delta I_{M}+\delta I_{G}=0
\end{equation}

However, this can only be true $iff$
\begin{equation}
\delta I_{M}+\delta I_{G}=0\iff \delta I_{G}=-\delta I_{M}%
\end{equation}

The curvature scalar $R$ can be defined as $g^{\mu \nu}R_{\mu \nu},$ and the
following standard equations are valid (\cite{SW-2}, P.364):%

\begin{equation}
\delta \left(  \sqrt{-g}R\right)  =\sqrt{-g}R_{\mu \nu}\delta g^{\mu \nu}%
+R\delta \sqrt{-g}+\sqrt{-g}g^{\mu \nu}\delta R_{\mu \nu}%
\end{equation}

\begin{equation}
\delta R_{\mu \nu}=(\delta \Gamma_{\mu \lambda}^{\lambda})_{;\nu}-(\delta
\Gamma_{\mu \nu}^{\lambda})_{;\lambda}%
\end{equation}

\begin{equation}
\sqrt{-g}g^{\mu \nu}\delta R_{\mu \nu}=\frac{\partial}{\partial x^{\nu}}%
(\sqrt{-g}g^{\mu \nu}\delta \Gamma_{\mu \lambda}^{\lambda})-\frac{\partial
}{\partial x^{\lambda}}(\sqrt{-g}g^{\mu \nu}\delta \Gamma_{\mu \nu}^{\lambda})
\end{equation}

\begin{equation}
\delta \sqrt{-g}=\frac{1}{2}\sqrt{-g}g^{\mu \nu}\delta g_{\mu \nu}%
\end{equation}

\begin{equation}
\delta g^{\mu \nu}=-g^{\mu \rho}g^{\nu \sigma}\delta g_{\rho \sigma}%
\end{equation}

Eq.(128) vanishes when we integrate over all space (\cite{SW-2}, P. 364).
Thus, for the pure gravitational part,we have%
\begin{align}
\delta I_{G}  & =\frac{1}{16\pi G_{N}}\int \sqrt{-g}\times \nonumber \\
& \left[  R_{\mu \nu}g^{\mu \rho}g^{\nu \sigma}\delta g_{\rho \sigma}-\frac{1}%
{2}g^{\mu \nu}\text{ }R\text{ }\delta g_{\mu \nu}\right]  \text{ }%
d^{4}x\nonumber \\
\delta I_{G}  & =\frac{1}{16\pi G_{N}}\int \sqrt{-g}\left[  R^{\mu \nu}-\frac
{1}{2}g^{\mu \nu}\text{ }R\right]  \delta g_{\mu \nu}\text{ }d^{4}x
\end{align}

From Eqs. (121), (125) and (131), we have%
\begin{align}
\delta I_{G}  & =-\delta I_{M}\Longrightarrow \frac{1}{16\pi G_{N}}\left[
R^{\mu \nu}-\frac{1}{2}g^{\mu \nu}\text{ }R\right]  =\nonumber \\
& -\frac{1}{2}T^{\mu \nu}\nonumber \\
\delta I_{G}+\delta I_{M}  & =R^{\mu \nu}-\frac{1}{2}g^{\mu \nu}\text{ }R+8\pi
G_{N}T^{\mu \nu}=0
\end{align}

By using%
\begin{equation}
g_{\alpha \gamma}g_{\beta \delta}A^{\gamma \delta}=A_{\alpha \beta}%
\end{equation}

and redefining the resulting indices as $\mu$ and $\nu$, we get%
\begin{equation}
R_{\mu \nu}-\frac{1}{2}g_{\mu \nu}\text{ }R=-8\pi G_{N}T_{\mu \nu}%
\end{equation}

It should be noted that all terms in the Eq.(134) are already present in the
Eqs.(15) and (18), as such the underlying symmetry (general coordinate
invariance) of the Eq.(7) is still preserved in a covariant manner. Eq.(132)
ensures the conservation of energy-momentum (which is a statement of general
covariance \cite{SW-2}, P. 361). Thus, the Weyl's action given in the Eq.(123)
would be stationary / invariant with respect to the variation in $g_{\mu \nu},$
$iff$ \ Eq.(132) holds. Interestingly, it holds because Eq.(132) is the
Einstein field equations, and hence the full Weyl's action is stationary with
respect to the variation in $g_{\mu \nu}.$\emph{This is precisely what we
expect: that the invariance of Weyl's action is maintained by inducing general
relativity. Hence the general covariance property of Eq.(7) has been revealed
because the statement that }$\delta I_{eff}$ \emph{should vanish is "generally
covariant", and this leads to the energy-momentum conservation }(\cite{SW-2},
P. 361).

Conclusively, the perturbative aspect of strong gravity theory (i.e. Eq.(27))
possesses quantum gauge theory (Eq.(106)) and gravity theory (Eq.(134)); thus
proving the existence of gauge-gravity duality in the strong gravity formulation.

\subsection{Ultraviolet Finiteness}

The strong gravity program adopts the Wilsonian viewpoint on quantum field
theory. Here the basic input data to be fixed \emph{ab initio }are the kind of
quantum fields (i.e., gluon fields) carrying the theory's degrees of freedom
(one graviton equals two gluons: BCJ construction), and the underlying
symmetry (spherical/rotational symmetry). The fact that two gluons are used to
construct spacetime metric means that the resulting gravity must be
point-like. This fact is encoded in the three-dimensional Dirac delta
functions in the first part of the Eqs.(19), and (30). The point-like nature
of gravity in this picture is the origin of ultraviolet (UV) divergence. The
question here is: Is Eq.(106) (the effective potential carried by Eqs.(27)) UV
finite, or perturbatively renormalizable? This question can be answered by
using Eqs.(106) and (108):%

\begin{equation}
V_{pert}(r)=\frac{F}{r}-\frac{4F}{3r}\left[  1-\frac{\beta_{1}}{r}+\frac
{\beta_{1}^{2}}{2r^{2}}-\frac{\beta_{1}^{3}}{6r^{3}}+\frac{\beta_{1}^{4}%
}{24r^{4}}-\text{ }...\right]
\end{equation}

It is to be noted, from \textbf{subsection C} of \textbf{section III}, that
the expression for $\beta_{1}$( with dimension of $GeV^{-1}\longrightarrow
E^{-1}$) contains inverse of \textbf{boson fields} dimension ($E^{-1}$),
and\textbf{\ fermion fields} dimension ($G_{f}^{3/2}\longrightarrow E^{-3/2}%
$). So it suffices to posit that $\beta_{1}$ contains both boson and fermion
fields: A perfect replica of supersymmetric fermion-boson field duality. Let
us now test for the UV behavior of the Eq. (106):%
\begin{equation}
\underset{r\longrightarrow0}{V_{pert}(r)}\text{ }=\infty-\infty \left[
1-\infty+\infty-\infty+\infty-\text{ }...\right]  =0
\end{equation}

Clearly Eq.(136) is a host of infinities, but they all cancel out, thus
rendering Eq.(106) UV finite. Hence, strong gravity theory has \textbf{UV
regularity}. Interestingly, this is the main conclusion of the theories of
supergravity ("enhanced cancellations").

\subsection{Breaking of Chiral Symmetry in Strong Gravity Theory}

QCD admits a \textbf{chiral symmetry} in the advent of vanishing quark masses.
\textbf{This symmetry is broken spontaneously by dynamical chiral symmetry};
and \textbf{broken explicitly by quark masses}. The nonperturbative scale of
dynamical chiral symmetry breaking is around $\Lambda_{x}\approx1GeV$
\cite{TOGO26}. Apparently, the chiral symmetry in the strong gravity is broken
spontaneously by its inherent dynamical chiral symmetry breaking $G_{f}%
^{-1}=\Lambda_{QCD}=\Lambda_{x}\approx1GeV$. In much the same spirit, the
calculated value of mass scale of the theory reverberates the existence of the
approximate symmetry in the strong interaction: $m=1.29GeV$ and $G_{f}%
^{-1}=\Lambda_{QCD}\approx1GeV$.

\section{Confinement and Asymptotic Freedom}

In the past few decades it became a common knowledge that confinement is due
to a linearly rising potential between static test quarks / gluons in the
4-dimensional pure Yang-Mills theory (see Eq.(105)). The fact that confinement
(i.e. non-perturbative aspect of QCD) is a simple consequence of the strong
coupling expansion means that an infinitely rising linear potential becomes
highly non-trivial in the weak coupling limit of the theory. This short-scale
weak coupling limit is called asymptotic freedom \cite{MCORN,CGATT}. By all
standards, these two properties of QCD contradict all previous experience in
physics with strong force decreasing with distance. The asymptotic freedom
part of the paradox has been correctly resolved \cite{TOGO10,TOGO11}, leaving
out the hitherto unresolved color confinement property of the non-perturbative
QCD regime. As we have remarked previously, a complete theory of strong
interaction should be able to explain these two properties of QCD
simultaneously (i.e., the dominance of asymptotic freedom at the small scale
distances (quark-gluon regime) and the emergence of infrared slavery
(confinement) at long scale distances (hadronic regime)). These dual
properties of QCD are succinctly depicted in the Eq.(109).

The linearly rising potential means that the potential between a static
gluon-gluon pair keeps rising linearly as one tries to pull the two
constituents apart (see Eq.(105)). Thus they are confined in a strongly bound
state \cite{MCORN}. Based on the dynamics of Eq.(105), an infinite amount of
energy would be required to pull the two constituents of bound glueball/meson
state apart.

The resulting force of strong gravity theory is called Yang-Mills-Gravity
force ($F_{YMG}(r)$), because Eq.(7) $-$which gives rise to the confining
potential $-$ is the Weyl's action for \textbf{gravity }(\cite{ASCS},
P.322)\textbf{, }and the action in the Eq.(27)\textbf{\ }$-$ which gives rise
to the perturbative QYMT $-$ also contains Einstein-Hilbert action for
\textbf{gravity. }To explain the behavior of this force at both small and
large distance scales, we differentiate Eq.(109) with respect to the
gluon-gluon separating distance $r$ (and taking into consideration Eq.(107)):%
\begin{equation}
F_{YMG}(r)=-\frac{F\left(  1-C_{F}\text{ }e^{-\beta_{1}/r}\right)  }{r^{2}%
}-\frac{FC_{F}\beta_{1}e^{-\beta_{1}/r}}{r^{3}}+\sigma
\end{equation}

The summing graphs of strong gravitational gluodynamics are shown in the
\textbf{Fig.2. }The blue graphs are the graphs of the effective pure
Yang-Mills potential (Eq.(109)), while the red plots are the graphs of the
Yang-Mills-Gravity force (Eq.(137)). It is easy to show that these equations
possess UV asymptotic freedom (albeit with tamable infinities) and infrared
(IR) slavery behaviors of the QCD. For us to see these behaviors, the
following facts are in order: (i) If the radial derivative of potential is
positive, then the force is attractive. (ii) If the radial derivative of
potential is negative, then the force becomes repulsive \cite{HJW}. (iii)
Since only color singlet states (hadrons)/ or dressed glueball can exist as
free observable particles, we multiplied the gluon-distance scale (in the
\textbf{Figs.2} and \textbf{3}) by factor of 10 in order to convert gluon
radius to the more observable hadronic radius (in line with the Eq.(87)). (iv)
The graphs in the \textbf{Figs.2} and \textbf{3} are plotted by using the
highly interactive plotting software \cite{DAVID}.

The strong interaction is observable in two areas: (i) on a shorter distance
scale ( for $10^{-19}GeV^{-1}\leq r\leq3.0277GeV^{-1}$ ), $F_{YMG}(r)$ is
repulsive (i.e., negative force) and reducing in strength as we probe shorter
and shorter distances (up to Planck length ($10^{-19}GeV^{-1}$)). This makes
Eq.(137) to be compatible with the asymptotic freedom property of QCD, where
the force that holds the quark-antiquark or gluon-gluon together decreases as
the distance between them decreases. Being a repulsive force (within the range
$10^{-19}GeV^{-1}\leq r\leq3.0277GeV^{-1}$ ), it would disallow the formation
of quark-antiquark / gluon-gluon singularity because the constituents can only
come close up to a minimum distance scale at which the repulsive force would
be strong enough to prevent further reduction in their separating distance.
(ii) On a longer distance scale ($r\geq3.0278GeV^{-1}$), $F_{YMG}(r)$ becomes
attractive (i.e., positive force). Here $F_{YMG}(r)$ does not diminish with
increasing distance. After a limiting distance ($r=10^{4}GeV^{-1}$) has been
reached, it remains constant at a strength of $0.299GeV^{2}$ (no matter how
much farther the separating distance between the quarks /gluons). Meanwhile,
the linearly rising potential keeps on increasing \emph{ad infinitum (see the
blue curve in the Fig.3).} This phenomenon is called color confinement in QCD.
The explanation is that the amount of workdone against a force of
$0.299GeV^{2}$ ($=2.449\times10^{5}N$) is enough to create
particle-antiparticle pairs within a short distance $r=10^{4}GeV^{-1}%
=1.972\times10^{-12}m$ than to keep on increasing the color force indefinitely.

By using \cite{DAVID}, we demonstrate that Eqs.(109) and (137) are consistent
and well-behaved down to the Planck scale:\ (i) At $r=10^{-19}GeV^{-1}%
=1.972\times10^{-35}m$ (Planck length), $V_{YM}^{eff}(r)=2.58\times10^{19}GeV$
(Planck energy) and $F_{YMG}(r)=-2.6\times10^{38}GeV^{2}.$ The negative sign
of $F_{YMG}(r)$ is the hallmark of the asymptotic freedom and the weakness of
gravitational field ($F_{YMG}(r)<0$) at the Planck scale! This would also
disallow the formation of singularity at the centre of a blackhole (see
\textbf{Fig.3} for more details). Based on the foregoing, we therefore assert
that strong gravity theory is consistent and well-behaved down to Planck
distance scale ($\sim10^{-19}GeV^{-1}$) .

\begin{figure}[ht!]
\begin{center}
\psfig{file=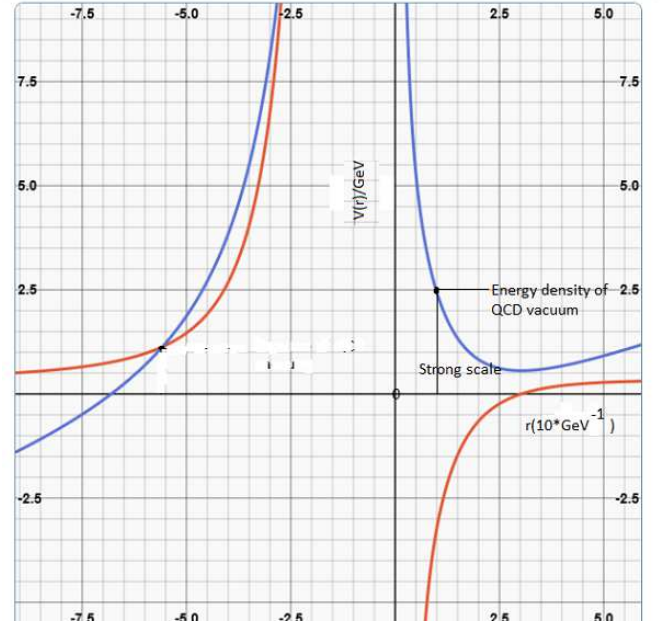,scale=0.55,angle=0,clip=}
\end{center}
\caption{Summing graphs of strong gravitational gluodynamics.}
\end{figure}%

\begin{figure}
\begin{center}
\psfig{file=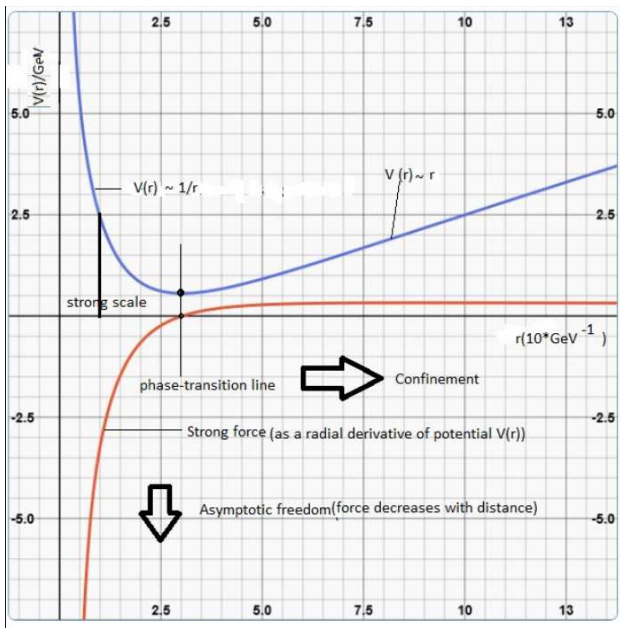,scale=0.45,angle=0,clip=}
\end{center}
\caption
{Graphs of pure Yang-Mills potential (in blue) and Yang-Mills-Gravity force (in red).}
\end{figure}%

\subsection{Energy density of QCD vacuum}

The scale invariance of the strong gravity is broken at $\Lambda_{QCD}%
\approx1GeV$ (\cite{ASCS}. P. 324). Hence the associated distance scale would
be given as $r_{g}=G_{f}=1GeV^{-1}.$ In terms of the observable hadronic
radius (see Eq.(87)), we have $r_{h}=10GeV^{-1}=1.972\times10^{-15}m=1.972$
$fm.$ The QCD potential at this distance scale is given as $V_{YM}%
^{eff}=2.495761GeV$ from the \textbf{Fig.3}, and the energy density
($\varepsilon$) of the QCD vacuum is calculated as:%
\begin{equation}
\varepsilon=\frac{V_{YM}^{eff}}{(r_{h})^{3}}=\frac{2.495761GeV}{(1.972)^{3}%
fm^{3}}=0.325GeV\text{ }/\text{ }fm^{3}%
\end{equation}

Eq.(138) is to be compared with the value calculated from the Lattice QCD
($\varepsilon \approx0.33GeV$ $/$ $fm^{3}$) (\cite{HENG}, P.54).

\section{ Existence of Quantum Yang-Mills Theory on R$^{4}$}

The existence of quantum Yang-Mills theory on $R^{4}$ (with its characteristic
mass gap) is one of the seven (now six) Millennium prize problems in
mathematics that was put forward by Clay Mathematics Institute in 2000
\cite{TOGO27}. The problem is stated as follows:

Prove that for any compact simple gauge group $G=SU(N)$, a fully renormalized
quantum Yang-Mills theory exists on $R^{4}$ and has a non-vanishing mass gap.

\subsection{Solution-plan}

The first thing to note here is that Yang-Mills theory is a non-abelian gauge
theory, and the idea of a gauge theory emerged from the work of Hermann Weyl
\cite{TOGO27A} (the same Weyl that formulated the Weyl's action that was used
in the formulation of strong gravity theory, based on the
\emph{Weyl-Salam-Sivaram's approach }\cite{ASCS}).

The Maxwell's theory of electromagnetism is one of the classical examples of
gauge theory. In this case, the gauge symmetry group of the theory is the
abelian group $U(1)$. If $A$ designates the $U(1)$ gauge connection (locally a
one-form on spacetime), then the potential of the field is the \textbf{linear}
two-form $F=dA$. To formulate the classical version of the Yang-Mills theory,
we must replace the gauge group $U(1)$ of electromagnetism by a compact gauge
group $SU(N)$, and the potential arising from the field would be a generalized
form of the Maxwell's: $F=dA+A\Lambda A $. This formula still holds at the
quantum level of the theory because Yang-Mills field shows quantum behavior
that is very similar to its classical behavior at short distance scales
(\cite{TOGO27}, P.1-2). However, the Maxwell's theory must be replaced by its
quantum version (i.e. QED; photon-electron interaction), and the nonlinear
part ($A\Lambda A$) must now describe the self-interaction of gluons (which is
the source of nonlinearity of the theory). The fact that the physics of strong
interaction is described by a non-abelian gauge group $G=SU(3)$ (i.e. QCD),
suggests immediately that the potentials of the four-dimensional quantum
Yang-Mills field must be the sum of the linear QED ($dA$) and nonlinear QCD
($A\Lambda A$) potentials at quantum level. Thus the first composite hurdle
for any would-be solution of the problem to cross is to: (1) obtain $QED+QCD$
potential at short distances with a single unified coupling constant. (2) The
two potentials must perfectly explain the individual physics of QED and QCD at
the quantum scale. (3) The two potentials must be obtained from
a\textbf{\ four-dimensional quantum gauge theory. }To surmount this composite
hurdle, one must first of all\emph{\ establish the existence of
four-dimensional quantum gauge theory with gauge group }$G=SU(N)$, and then
every other thing will follow naturally.

\subsubsection{Jaffe-Witten Existence Theorem (\cite{TOGO27},P.6)}

The official description of this (i.e. Yang-Mills existence and mass gap)
problem was put forward by Arthur Jaffe and Edward Witten. Their
\emph{existence theorem} is briefly paraphrased as follows: The existence of
four-dimensional quantum gauge theory (with gauge group $SU(N)$) can be
established mathematically, by defining a quantum field theory with local
quantum field operators in connection with the local gauge-invariant
polynomials, in the curvature $F$ and its covariant derivatives, such as
$TrF_{ij}F_{kl}(x)$. In this case, the correlation functions of the quantum
field operators should be in agreement with the predictions of
\textbf{perturbative renormalization (i.e. the theory must have UV
regularity)} and \textbf{asymptotic freedom (i.e. the weakness of strong force
at extremely short-distance scale)}; and there must exist a stress tensor and
an operator product expansion, admitting well-defined local singularities
predicted by asymptotic freedom.

By using the \emph{eye of differential geometry}, we observed that the
solution to the problem is concealed in the mathematical structures rooted in
the differential geometry . In other words, the above-stated existence theorem
is the mathematical description of the \emph{strong gravity formulation}.

\subsubsection{R$^{4}-$Weyl-Salam-Sivaram Theorem\cite{ASCS}}

The Weyl-Salam-Sivaram theorem is in fact the geometrical interpretation of
the Jaffe-Witten existence theorem. In the following, the local quantum field
operators are the two strong tensor fields ($G_{\mu}^{a}(x)$ and $G_{\nu}%
^{b}(x);$ two gluons forming double-copy construction) used to construct the
spacetime metric in the \textbf{section III} of this paper. These local
quantum fields have a direct connection (via $g=\det(G_{\mu}^{a}G_{\nu}%
^{b}\eta_{ab})$) with the gauge-invariant \ local polynomials in the curvature
$C$ and its covariant derivatives: $\sqrt{-g}C_{\alpha \beta \gamma \delta
}C^{\alpha \beta \gamma \delta}(x)$. Note that "$Tr$" in the Jaffe-Witten
existence theorem denotes an invariant quadratic form on the Lie algebra of
group $G$. Similarly, $\sqrt{-g}$ in the Weyl-Salam-Sivaram theorem denotes an
invariant quadratic form on the gauge group $SU(3).$ The correlation function
in this case is nothing but the spacetime metric ($g_{\mu \nu}(x)$) constructed
out of the two local quantum fields ($G_{\mu}^{a}(x)$ and $G_{\nu}^{b}(x)$),
and used as a function of the spatial cum temporal distance between these two
random variables (gluons). We have painstakingly demonstrated that this
spacetime metric agrees, at short distance scales, with the predictions of
asymptotic freedom (i.e. the weakness of strong force at extremely short
distance scales (see \textbf{section IX})) and perturbative renormalization
(i.e. the existence of UV regularity of the theory at short distances; the
theory should be able to regularize its own divergences at extremely short
distance scales, say, $r=0$ (see \textbf{subsection C of section VIII})).
There also exist a stress energy-momentum tensor (\textbf{Eq.(17)}), and field
product expansion (\textbf{Eq.(18)}), having local singularities encoded in
the three-dimensional Dirac delta functions (\textbf{Eqs. (19) and (30)})
predicted by asymptotic freedom. \textbf{Overall the broken-scale-invariant
Weyl action (Eq.(27)) is the required perturbative four-dimensional quantum
gauge field theory with its inherent gauge group }$SU(3)$\textbf{\ that gives
rise to color/Casimir factor }$4/3$\textbf{\ (Eq.(107))}. However, for this
statement to be valid the theory must possess both\emph{\ QED} and\emph{\ QCD}
\emph{potentials (i.e. }$F=dA+A\Lambda A$\emph{). Happily, the theory does
possess these potentials with a single coupling constant (see Eqs. (106),
(110), (111) and (113)).}

The fact that the scale invariance of Weyl action is broken at the strong
scale $\Lambda_{QCD}=G_{f}^{-1}\approx1GeV$ (\cite{ASCS}, P.324) $-$ which is
equal to its dynamical chiral symmetry breaking scale \cite{TOGO26} $-$ is a
clear indication of the existence of proton as the fundamental hadron of the
theory. In this case, one must therefore investigate the ground state (neutron
state) of the proton state using isospin symmetry. But for this to be
possible, the gauge group that describes isospin symmetry must exist within
the framework of the theory. This is where custodial symmetry (Eq.(77)) kicks
in. The vector subgroup of custodial symmetry is in fact the isospin symmetry:
$SU(2)_{L}\times SU(2)_{R}\longrightarrow SU(2)_{V}$ \cite{TOGO28}. This
isospin symmetry then demands that the Hamiltonian ($H$) of proton-neutron
state must be zero. However, the near mass-degeneracy of the neutron and
proton in the $SU(2)$ doublet representation points to an approximate isospin
symmetry of the Hamiltonian describing the strong interaction \cite{DGR,
CITZ}. The mass gap in this picture is nothing but the energy difference
between the two sub-states of the proton-neutron configuration: $m_{gap}%
=m_{n}-m_{p}\approx1.29MeV.$ Hence the mass formula of QCD (Eq.(93)) and the
stable Higgs boson mass (see next section) must be expressed in terms of this
mass gap.

Conclusively, the two gauge groups that are needed to accurately describe the
solution to this Millennium prize problem are $\ SU(3)-$ for the establishment
of the existence theorem$-$ and $SU(2)-$ for describing the mass gap of the
solution.\  \textbf{Hence, the Weyl-Salam-Sivaram existence theorem of strong
gravity puts quantum gauge field theory (QFT) on a solid mathematical footing
of the differential geometry; in this sense, QFT is a full-fledged part of
mathematics.}

\section{Stability of Vacuum: A hint for Planck scale physics from
$m_{H}=126GeV$}

The 126GeV Higgs mass seems to be a rather special value, from all the \emph{a
priori} possible values, because it just at the edge of the mass range
implying the stability of Minkowski vacuum all the way down to the Planck
scale \cite{EFEMI3}. If one uses the Planck energy ($G_{N}^{-1}\approx
10^{19}GeV$) as the cutoff scale, then the vacuum stability bound on the mass
of the Higgs boson is found to be 129GeV. That is, vacuum stability requires
the Higgs boson mass to be $m_{H}=129GeV$ \cite{EFEMI4} . A new physics beyond
SM is thus needed to reconcile the discrepancy between 126GeV and 129GeV mass
of Higgs boson. The first thing to observe here is that the vacuum stability
bound on the mass of Higgs boson ($m_{H}=129GeV$) has exactly the same
"number-structure" with the values that we have been working with in this paper.

By using Eq.(93), we can write%
\begin{equation}
m_{H}=const.\times m
\end{equation}

Comparing the energy scale of the pure Yang-Mill propagator in the Eq.(29)
($k^{-2}=1\times10^{17}GeV$) with the Planck scale ($\approx G_{N}^{-1}%
\approx10^{19}GeV$) shows a magnitude difference of $10^{2}.$ By using this
value as our constant (i.e. $const.=\frac{G_{N}^{-1}}{k^{-2}}$), we get
exactly $m_{H}=129GeV$:%
\begin{equation}
m_{H}=m\left(  \frac{G_{N}^{-1}}{k^{-2}}\right)  =129GeV
\end{equation}

Eq.(140) is very important because: (1) it shows the coupling of Higgs mass
($m_{H}$) to the fundamental mass, and mass gap of the QCD vacuum
($m=1290MeV=10^{3}\times m_{gap}$). (2) It connects Higgs mass to the Planck
energy scale.\textbf{\ }To show the vacuum stability property of the Eq.(140),
we eliminate the fundamental mass of the QCD vacuum by using the value of
critical temperature from Eq.(89) ($T\equiv m=10T_{c}$):%
\begin{equation}
m_{H}=T\left(  \frac{G_{N}^{-1}}{k^{-2}}\right)  =129GeV
\end{equation}

Obviously, $T>T_{c}$ (see subsection A of section VII). This is the well-known
vacuum stability condition in the \textbf{second-order phase transition
theory}; while the condition for vacuum instability is $T<T_{c}$ (see
\cite{FRAG} and the references therein).

The mass range of the Higgs boson that would allow the stability of vacuum is
given as \cite{IATO}:%
\begin{equation}
123GeV\leq m_{H}\leq129GeV
\end{equation}

By taking the average value of Eq.(142), we have
\begin{equation}
m_{H}^{avg}=\frac{123GeV+129GeV}{2}=126GeV
\end{equation}

Clearly, 126GeV Higgs mass is special because it just at the midpoint of the
mass range that guarantees the stability of the vacuum.

\section{The Enigmatic Neutrino}

\begin{quotation}
"A cosmic mystery of immense proportions, once seemingly on the verge of
solution, has deepened and left astronomers and astrophysicists more baffled
than ever. The crux ...is that the vast majority of the mass of the universe
seems to be missing." - \textbf{William J. Broad (1984)}

"A billion neutrinos go swimming in heavy water: \ one gets wet." -
\textbf{Michael Kamakana}
\end{quotation}

Studying the properties of neutrinos has been one of the most exciting and
challenging activities in particle physics and astrophysics ever since Pauli,
"the unwilling father" of neutrino, proposed their existence in 1930 in order
to find the desperate remedy for the law of conservation of energy, which
appeared to be violated in $\beta-$decay processes. Since then, many hidden
facts about neutrinos have been unveiled step by step\cite{FBP, SWJ}. In spite
of their \emph{weakly interacting nature}, we have so far gathered an
avalanche of knowledge about neutrinos. From the neutrino oscillation
experiments (an effort that has been duly awarded the 2015 Nobel prize in
physics \cite{TKAB}), we learned that there are two major problems that plague
neutrino physics:

\textbf{(1)} Determination of the absolute masses of neutrinos. The results
from the neutrino oscillation experiments have confirmed the massive nature of
neutrino. However, this confirmation provides a crack in the foundation of the
Standard Model (SM) of particle physics, because SM treats neutrinos as
massless particles. This disagreement between SM and experimental results
(which opens a new door to the physics beyond SM) constitutes what is called
"neutrino mass problem"\cite{TKAA7, ABMD3, ASNE,KEG,PAD,YABD,FANE,JAHN}.

\textbf{(2) }Another major problem in the neutrino physics that is somehow
related to the one above-mentioned, is to establish whether the neutrinos with
definite masses $m_{k}$ are Dirac particles (with particles and antiparticles
being different objects thereby conserving the lepton number) or Majorana
particles (with particles and antiparticles being the same thereby violating
lepton number). An experimental distinction between these two seems to be much
more complicated than the confirmation of non-vanishing mass of the neutrino.
These are the two major problems in neutrino physics that have hitherto defied
all solutions.

Based on the formulation of the strong gravity theory (that hadronic
interactions become weak in strength at small invariant separation), we assert
that the absolute masses of neutrinos are actually calculable. More
importantly, we will demonstrate,\textbf{\ in this section}, that neutrinos
are Majorana particles! In few lines, we explain the theoretical properties of
neutrino's nature that form the basis for using the strong gravity formulation.

\textbf{(i) }All types of neutrino participate in weak nuclear and
gravitational interactions with ordinary matter \cite{YVKO}. This means that
their physics can be explained by using a gas of weakly coupled particles
system (a configuration that we used to solve the problem of asymptotic
freedom (i.e., calculation of the dimensionless strong coupling constant at
the starting point of QCD evolution in this paper)). The fact that strong
gravity combines strong nuclear force (which becomes weak at extremely short
distance scale: $r\ll \Lambda_{QCD}^{-1}$) and gravitational force into one
unified force makes the determination of neutrino masses possible within the
framework of massive spin-2 field theory with $D=5$: note that the Lagrangian
of the Majorana neutrino is valid only when $D=5$.

\textbf{(ii)} Majorana neutrino Lagrangian possesses symmetry axis /
CP-symmetry (\cite{TOGO19}, P. 203-205).

These two points form the basis of our solution-plan for solving the neutrino
mass problem. This approach shows a compelling interplay between gravitation
and principle of linear superposition of different mass eigenstates of
neutrino as alluded to in \cite{DVACB}.

\subsection{Effective Majorana Mass Matrix}

Since the Majorana neutrino has only left-handed chiral field $\nu_{L},$ which
is present in the SM, it is therefore natural to ask if it possible for SM
neutrinos to have Majorana masses. The simple answer is that it is not
possible, due to the fact that the left-handed chiral field $\nu_{L}$ has weak
\textbf{isospin triplet} with hypercharge $Y=-2.$ The fact that SM does not
contain any weak isospin triplet with $Y=2$ clearly \ shows that it is not
possible to have a renormalizable Lagrangian term which can generate Majorana
neutrino masses (\cite{TOGO19}, P. 205).

However, the lowest dimensional Lagrangian which could generate Majorana
neutrino masses that one can construct with the SM fields, respecting the SM
symmetries, is the lepton number violating Lagrangian (with $D>4$)
(\cite{TOGO19}, P. 216):%
\begin{equation}%
\mathcal{L}%
_{d}=M_{X}^{4-D}\underset{\alpha \beta}{\sum}g_{\alpha \beta}(L_{\alpha
L}^{^{\prime}T}\tau_{2}\Phi)C^{\dag}(\Phi^{T}\tau_{2}L_{\beta L}^{^{\prime}%
})+H.c.
\end{equation}

where $M_{X}$ is a heavy mass ( of a single color triplet Higgs scalar )
characteristic of the symmetry-breaking scale of the high-energy unified
theory, $D$ is called a \emph{dimension-D operator} and its value in this case
is $D=5.$ $g_{\alpha \beta}$ is a yet-unknown symmetric $3\times3$ matrix of
coupling constants. With $D=5,$ Eq.(144) becomes%
\begin{equation}%
\mathcal{L}%
_{5}=\frac{1}{M_{X}}\underset{\alpha \beta}{\sum}g_{\alpha \beta}(L_{\alpha
L}^{^{\prime}T}\tau_{2}\Phi)C^{\dag}(\Phi^{T}\tau_{2}L_{\beta L}^{^{\prime}%
})+H.c.
\end{equation}

The electroweak symmetry breaking VEV ($=\upsilon=246GeV$ \cite{DBAL}) of the
Higgs field leads to the Majorana neutrino mass term(\cite{TOGO19}, P. 216);%
\begin{equation}%
\mathcal{L}%
_{mass}^{M}=\frac{1}{2}\frac{\upsilon^{2}}{M_{X}}\underset{\alpha \beta}{\sum
}g_{\alpha \beta}\text{ }\nu_{\alpha L}^{^{\prime}T}C^{\dag}\nu_{\beta
L}^{^{\prime}}+H.c.
\end{equation}

From Eq.(146), \ the Majorana mass matrix has elements (\cite{TOGO19}, P. 216)%
\begin{equation}
M_{\alpha \beta}^{L}=\frac{\upsilon^{2}}{M_{X}}\text{ }g_{\alpha \beta}%
\end{equation}

with (\cite{TOGO19}, P. 208)%
\begin{equation}
M_{\alpha \beta}^{L}=M_{\beta \alpha}^{L}%
\end{equation}

Eq.(148) is the reason why the $g_{\alpha \beta}$ matrix must be symmetric.
With $\alpha=\beta=0,1,2$, Eq.(147) reduces to%
\begin{equation}
M_{00}^{L}=\frac{\upsilon^{2}}{M_{X}}\text{ }g_{00}%
\end{equation}

\begin{equation}
M_{11}^{L}=\frac{\upsilon^{2}}{M_{X}}\text{ }g_{11}%
\end{equation}

\begin{equation}
M_{22}^{L}=\frac{\upsilon^{2}}{M_{X}}\text{ }g_{22}%
\end{equation}

\emph{(It is worth noting that if all the diagonal elements of }%
$g_{\alpha \beta}$\emph{\ are all 1's, then the Eqs. (147) and 149-151 reduce
to Eq.(74).)}

The gravitational potential ($g_{\mu \nu}$) which is capable of representing a
combined gravitational and electromagnetic field outside a \textbf{spherically
symmetric material distribution }is given as \cite{MIW};%
\begin{equation}
g_{\mu \nu}=\left(
\begin{array}
[c]{cccc}%
g_{00} & g_{01} & 0 & 0\\
g_{10} & g_{11} & 0 & 0\\
0 & 0 & g_{22} & 0\\
0 & 0 & 0 & g_{33}%
\end{array}
\right)
\end{equation}

where%
\begin{equation}
g_{00}=\frac{(1-\frac{m}{2r})^{2}}{(1+\frac{m}{2r})^{2}}+\frac{\zeta^{2}%
}{r(1+\frac{m}{2r})^{2}}%
\end{equation}

\begin{equation}
g_{01}=g_{10}=-\frac{\zeta(1+\frac{m}{2r})}{r^{1/2}}%
\end{equation}

\begin{equation}
g_{11}=(1+\frac{m}{2r})^{4}%
\end{equation}

\begin{equation}
g_{22}=g_{11}r^{2}%
\end{equation}

\begin{equation}
g_{33}=g_{22}\sin^{2}\theta=g_{11}r^{2}\sin^{2}\theta
\end{equation}

The quantity $m$ represents an effective gravitational mass, and $\zeta$ is an
electric-charge dependent parameter \cite{MIW}. Since neutrinos are
electrically neutral, we set $\zeta$ to zero: $\zeta=0.$ Hence Eq.(152)
reduces to%
\begin{equation}
g_{\mu \nu}=\left(
\begin{array}
[c]{cccc}%
g_{00} & 0 & 0 & 0\\
0 & g_{11} & 0 & 0\\
0 & 0 & g_{22} & 0\\
0 & 0 & 0 & g_{33}%
\end{array}
\right)
\end{equation}

and%

\begin{equation}
g_{00}=\frac{(1-\frac{m}{2r})^{2}}{(1+\frac{m}{2r})^{2}}%
\end{equation}

\[
g_{01}=g_{10}=0
\]

This matrix (Eq.158) has Euclidean space signature $++++.$ It's worth noting
that for us to impose Lorentz signature on the above matrix, we must invoke
the Levi-Civita indicator \ on the matrix to account for the special
relativity in the limiting case, and to also transform the metric from 4
dimensions to 3+1 dimensions. It doesn't matter whether we insert the Lorentz
signature before or after solving the Eq.(158), due to the fact that it is a
diagonalized matrix \cite{FIMMW}.

The fact that Majorana neutrino Lagrangian preserves CP symmetry means that it
possesses symmetry axis ($\theta=0$). The reason why Majorana neutrino
Lagrangian preserves CP symmetry is that Majorana particles are invariant to
CP transformation (because Majorana particle = Majorana antiparticle)
(\cite{TOGO19}, P. 203-205).

Consequently (by setting $\theta=0$), Eqs.(157-158) reduce to%
\begin{equation}
g_{33}=0
\end{equation}

\begin{equation}
g_{\mu \nu}=\left(
\begin{array}
[c]{cccc}%
g_{00} & 0 & 0 & 0\\
0 & g_{11} & 0 & 0\\
0 & 0 & g_{22} & 0\\
0 & 0 & 0 & 0
\end{array}
\right)
\end{equation}

Hence, Eq.(147) becomes%
\begin{equation}
M_{\alpha \beta}^{L}=\frac{\upsilon^{2}}{M_{X}}\text{ }\left(
\begin{array}
[c]{cccc}%
g_{00} & 0 & 0 & 0\\
0 & g_{11} & 0 & 0\\
0 & 0 & g_{22} & 0\\
0 & 0 & 0 & 0
\end{array}
\right)
\end{equation}

By solving Eq.(159) completely for mass $m$, we have%
\begin{equation}
m=\frac{2r(1-g_{00}^{1/2})}{(1+g_{00}^{1/2})}%
\end{equation}

Multiplying Eq.(159) by $\frac{(1+\frac{m}{2r})^{2}}{(1+\frac{m}{2r})^{2}}$
and solve the resulting equation completely for mass $m$;%
\begin{equation}
m=\pm2r[1-(g_{11}g_{00})^{1/2}]^{1/2}%
\end{equation}

where $\pm$ sign \ in Eq.(164) leads to the same result. By comparing Eq.(163)
with Eq.(164), we get%
\begin{equation}
g_{11}=\frac{1}{g_{00}}\left[  1-\frac{(1-g_{00}^{1/2})^{2}}{(1+g_{00}%
^{1/2})^{2}}\right]  ^{2}%
\end{equation}

Since our calculated value for $g_{00}$ is $g_{00}=0.1797,$ thus Eqs.(156) and
(165) reduce to%
\begin{equation}
g_{11}=3.8922
\end{equation}

\begin{equation}
g_{22}=3.8922r^{2}%
\end{equation}

We now look for an ingenious way to eliminate $r^{2}$ in Eq.(167). It is
tempting to straightforwardly use unit sphere formalism but this direct
approach will not work because $M_{\alpha \beta}^{L}$ is a linear superposition
of three different neutrino masses, albeit from the same source. The best
mathematical approach that we can use to circumvent this problem is the
3-sphere formulation (note that this approach is anchored on the fact that
3-sphere is a sphere in 4-dimensional Euclidean space) \cite{GEOLE, MAAPE}:%
\begin{align}
r^{2}  & =\overset{3}{\underset{i=0}{\sum}}(x_{i}-C_{i})^{2}=(x_{0}-C_{0}%
)^{2}+(x_{1}-C_{1})^{2}+\nonumber \\
& (x_{2}-C_{2})^{2}+(x_{3}-C_{3})^{2}%
\end{align}

We turn Eq.(168) on its head by using it to represent three spheres
(representing three types of neutrino) with common origin. This reduces
Eq.(168) to ordinary linear superposition of three spheres (in two-dimension,
they reduce to circles) with common origin / source. Suppose we further impose
the condition that the common origin is centred at zero (i.e., $x_{0}-C_{0}%
=0$), then Eq.(168) reduces to%
\begin{equation}
r^{2}=\overset{3}{\underset{i=1}{\sum}}(x_{i}-C_{i})^{2}=(x_{1}-C_{1}%
)^{2}+(x_{2}-C_{2})^{2}+(x_{3}-C_{3})^{2}%
\end{equation}

where $x_{1}-C_{1},$ $x_{2}-C_{2}$ and $x_{3}-C_{3}$ are the radii of the
spheres. By using unit sphere formalism individually on the three sphere,
Eq.(169) reduces to%
\begin{equation}
r^{2}=\overset{3}{\underset{i=1}{\sum}}(x_{i}-C_{i})^{2}=3
\end{equation}

Thus, Eq.(167) becomes%
\begin{equation}
g_{22}=11.6766
\end{equation}

and Eq.(161) reduces to%

\begin{equation}
g_{\mu \nu}=\left(
\begin{array}
[c]{cccc}%
0.1797 & 0 & 0 & 0\\
0 & 3.8922 & 0 & 0\\
0 & 0 & 11.6766 & 0\\
0 & 0 & 0 & 0
\end{array}
\right)
\end{equation}

With $M_{X}=1.63\times10^{16}GeV$ (see subsection C of section VI) and
$\upsilon=246GeV$ \cite{DBAL}, $\frac{\upsilon^{2}}{M_{X}}=3.7meV$ (see Eq.(79)).

Hence Eqs.(149-151) reduce to%
\begin{equation}
m_{0}=0.665meV
\end{equation}

\begin{equation}
m_{1}=14.401meV
\end{equation}

\begin{equation}
m_{2}=43.203meV
\end{equation}

where $m_{0}\equiv M_{00}^{L},$ $m_{1}\equiv M_{11}^{L},m_{2}\equiv M_{22}%
^{L}$ and $m_{3}\equiv0.$ And Eq.(162) reduces to%
\begin{equation}
M_{\alpha \beta}^{L}=3.7meV\left(
\begin{array}
[c]{cccc}%
0.1797 & 0 & 0 & 0\\
0 & 3.8922 & 0 & 0\\
0 & 0 & 11.6766 & 0\\
0 & 0 & 0 & 0
\end{array}
\right)
\end{equation}

For the purpose of book-keeping, we set $m_{0}\equiv m_{1},$ $m_{1}\equiv
m_{2}$ and $m_{2}\equiv m_{3}$. It is evident from Eqs.(173-175) that
$m_{1}<m_{2}<m_{3},$ which is clearly a Normal Mass Hierarchy signature. The
validity of which can also be confirmed by considering the approach of M.
Kadastik et al \cite{MKMR}.%
\begin{equation}
N_{1}=\frac{-m_{1}^{2}+m_{2}^{2}+3m_{3}^{2}}{2m_{1}^{2}+m_{2}^{2}}%
\end{equation}

with%
\begin{align}
N_{1}  & >1\rightarrow \text{ normal mass hierarchy}\nonumber \\
N_{1}  & <1\rightarrow \text{ inverted mass hierarchy}\nonumber \\
N_{1}  & \approx1\rightarrow \text{ degenerate masses}%
\end{align}

Taking the values of $m_{1},m_{2},$ and $m_{3}$ from Eqs.(173-175), Eq.(177)
gives the value $N_{1}\approx28,$ which satisfies the criterion of normal mass
hierarchy in Eq.(178).

The \ mass-squared difference is defined mathematically as%
\begin{equation}
\Delta m_{ij}^{2}=m_{i}^{2}-m_{j}^{2}%
\end{equation}

where $i>j.$ Based on the Eq.(179) (and taking into account Eqs.(173-175)), we
have the following equations:%
\begin{equation}
\Delta m_{21}^{2}=m_{2}^{2}-m_{1}^{2}=2.06\times10^{-4}eV^{2}%
\end{equation}

\begin{equation}
\Delta m_{31}^{2}=m_{3}^{2}-m_{1}^{2}=1.87\times10^{-3}eV^{2}%
\end{equation}

\begin{equation}
\Delta m_{32}^{2}=m_{3}^{2}-m_{2}^{2}=1.57\times10^{-3}eV^{2}%
\end{equation}

\subsubsection{Experimental Test}

\textbf{(1)} \ The combined results of all solar experiments with
Super-Kamiokande-I zenith spectrum and KamLAND data give $\Delta m_{sol}%
^{2}=\Delta m_{21}^{2}=2\times10^{-4}eV^{2}$ at $99.73\%$ C. L.\cite{HNWJ}.
This experimental value is compatible with our Eq.(180): Thus confirming the
validity of our $m_{1}$ and $m_{2}$ values.

\textbf{(2) }From the atmospheric neutrino oscillation experiments, the bound
on the mass of the heaviest neutrino is $m_{3}\gtrsim40meV$ \cite{HVKLK}. This
value experimentally confirms our value in the Eq.(175). We therefore assert
that the values of our $m_{1},m_{2}$ and $m_{3}$ conform with the experimental data.

\subsection{Observational Test}

The energy density of light massive neutrinos is given as (\cite{TOGO19}, P.
590-591):%
\begin{equation}
\Omega_{\nu}^{0}h^{2}=\frac{\overset{3}{\underset{i=1}{\sum}}m_{i}}{94.14eV}%
\end{equation}

where $\Omega_{\nu}^{0}h^{2}$ is the neutrino energy density (which is also
known as the \emph{Gershtein-Zeldovich limt or Cowsik-McClelland limit}) and
$\overset{3}{\underset{i=1}{\sum}}m_{i}$ is the sum of the three active
neutrino masses. From Eqs.(173-175), $\overset{3}{\underset{i=1}{\sum}}%
m_{i}=0.058269eV.$ To obtain an accurate result, we must convert the
calculated value of sum of the neutrino into two decimal places in conformity
with the denorminator of Eq.(183). Thus%

\begin{equation}
\overset{3}{\underset{i=1}{\sum}}m_{i}\approx0.06eV
\end{equation}

Consequently,
\begin{equation}
\Omega_{\nu}^{0}h^{2}\approx0.00064
\end{equation}

Eqs.(184) and (185) are the fiducial parameter values that have been taken to
be valid for the background Cosmology to be consistent with the most recent
cosmological measurements \cite{JBDH}. Here, it turns out that the South Pole
Telescope (SPT) cluster abundance is lower than preferred by either the WMAP9
or Planck+WMAP9 polarization data for the Planck base $\Lambda CDM$ model; but
assuming a normal mass hierarchy for the sum of of the neutrino masses with
$\sum m_{\nu}\approx0.06eV$ (\cite{TOGO18}, P.237 \& 239) the data sets are
found to be consistent at the $1.0\sigma$ level for WMAP9 and $1.5\sigma$
level for Planck+WMAP9 \cite{SBDCQ}. Obviously, our calculations confirm that
the Planck base $\Lambda CDM$ model's prediction of sum of the neutrino masses
is correct.

\section{Dark Energy}

For the strong gravity theory to be a complete theory of \textbf{QCD} and
\textbf{gravity}, it must be tested okay at both small and large scale
distances. The large distance ,here, is the cosmological scale where "dark
energy" is dominant $-$ Dark energy ($\rho$) is an unknown form of energy,
which was invented to account for the acceleration of the expanding universe.
The observed value (upper limit) of $\rho$ is $\rho^{observed}\approx
(2.42\times10^{-3}eV)^{4}$ \cite{AGRI} $-$ A major outstanding problem is that
most quantum field theories \textbf{naively} predict a huge value for the dark
energy: the prediction is wrong by a factor of $10^{120}$ \cite{AGRI1}. The
origin of the problem is now clear to us: \emph{Eqs.(118) and (120) clearly
show that the energy-momentum tensor is related to the invariant (Weyl)
Lagrangian density, but not to the total energy density of a vacuum, which is
not operationally measurable, due to quantum fluctuations!}

The energy density of any given system, such as the universe, is categorized
into two parts: one is due to the true vacuum ($\rho$) and the other to the
matter and radiation (pressure ($p$)) present in the system. These two types
of energy density are related by the energy-momentum tensor $T_{\mu \nu}$
\cite{SW-2}:%
\begin{equation}
T_{\mu \nu}=\text{ }\left(
\begin{array}
[c]{cccc}%
\rho & 0 & 0 & 0\\
0 & -p & 0 & 0\\
0 & 0 & -p & 0\\
0 & 0 & 0 & -p
\end{array}
\right)
\end{equation}

\emph{( Note that by putting Eq.(186) into Eq.(134), two things happen: (1)
The energy density of the true vacuum becomes negative, meaning repulsive
gravity and (2) the energy density of matter and radiation becomes positive,
meaning attractive gravity. These two results are compatible with the
observations. The gravity of ordinary matter/energy is always attractive,
while the gravity of true vacuum (i.e., dark energy) is always repulsive.)}

Since it has been observationally confirmed that the acceleration of the
expanding universe is controlled by the energy density of true vacuum ($\rho
$),but not by the matter/energy content of the universe, we can write (from
Eq.(186))%
\begin{equation}
T_{00}=\rho
\end{equation}

By combining Eq.(120) with Eq.(187), we get Eq.(50):%
\begin{equation}
\rho=g_{00}[E_{vac}]^{4}%
\end{equation}

Note that the Weyl Lagrangian density scales as $C_{\alpha \beta \gamma \delta
}C^{\alpha \beta \gamma \delta}\sim \lbrack E_{vac}]^{4}$ (where $E_{vac}$ denotes
the effective (Weyl) Lagrangian of the vacuum), due to the scale invariance of
the Weyl's action in the Eq.(7) (\cite{TOGO19}, P.206). To see the repulsive
nature of the dark energy, we combine Eqs.(134) and (188) to get
\begin{equation}
G_{\mu \nu}=-8\pi G_{N}\text{ }\left(
\begin{array}
[c]{cccc}%
\rho & 0 & 0 & 0\\
0 & 0 & 0 & 0\\
0 & 0 & 0 & 0\\
0 & 0 & 0 & 0
\end{array}
\right)
\end{equation}

where $G_{\mu \nu}\equiv R_{\mu \nu}-\frac{1}{2}g_{\mu \nu}$ $R$ is the
Einstein's tensor. The negative sign in the Eq.(189) is the hallmark of the
repulsive nature of dark energy. It is to be noted that we have not invoked
the presence of the famous cosmological constant ($\Lambda$) in the Eq.(134)
because it is not needed for the expanding or contracting universe
(\cite{TOGO18}, P.232). All what is needed for the accelerating expansion of
the universe, as currently observed, is the Eq.(189); while the combination of
the Eqs.(134) and (186) tell us that the universe will either expand (if the
right-hand side of the Eq.(134) is \textbf{negative (}$\rho$\textbf{)}) or
contract (if the right-hand side of the Eq.(134) is \textbf{positive (p)}).
Albert Einstein was right after all: the introduction of the fudged factor
($\Lambda$) was his greatest blunder. \emph{You cannot out-einstein Einstein!}

Using Eqs.(56) and (79), Eq.(188) becomes
\begin{equation}
\rho=(2.41\times10^{-3}eV)^{4}%
\end{equation}

Obviously, Eq.(190) compares favorably with the upper bound value of the
observed $\rho$ ($\rho^{observed}=(2.42\times10^{-3}eV)^{4}$) \cite{AGRI}.

It has long been suggested that the nonrelativistic massive neutrinos may give
a significant contribution to the energy density (i.e. the so-called dark
energy) of the universe (\cite{TOGO19}, P.590). This statement has been
confirmed to be true via Eqs.(176) and (188): with $E_{vac}=\upsilon^{2}%
/M_{X}=3.7meV$.

We understand, of course, that the energy of vacuum is extremely large (due to
quantum fluctuations) but the \textbf{strong gravity }and
\textbf{Majorana-neutrino Lagrangian (a conserved quantity that encodes the
information about the dynamics of the universe)} tell us that it is only the
effective Lagrangian of the universe that is physically measurable (i.e.
$\upsilon^{2}/M_{X}=3.7meV$).

\section{Dark Matter OR Leftover Yang-Mills-Gravity Force?}

\subsection{The Galaxy Rotation Problem (GRP)}

The GRP is the inconsistency between the theoretical prediction and the
observed galaxy rotation curves, assuming a centrally dominated mass
associated with the observed luminous material. The direct computation of mass
profiles of galaxies from the distribution of stars and gas in spirals and
mass-to-light ratios in the stellar disks, utterly disagree with the masses
derived from the observed rotation curves using Newtonian force law of
gravity. Based on the Newtonian dynamics, most of the mass of the galaxy had
to be in the galactic bulge near the center, and that stars and gas in the
disk portion should orbit the center at decreasing velocities with increasing
radial distance, away from the galactic center \cite{V.R,ECO,VTR}: This is
achieved by equating the centripetal force experienced by the orbiting
gas/stars to the Newton force law (\cite{TOGO18}, P.241):%
\begin{align}
F_{c}  & =F_{N},\nonumber \\
v  & =\sqrt{\frac{G_{N}\text{ }M}{r}}\implies v(r)\propto1/\sqrt{r}%
\end{align}

where $v$ is the speed of the orbiting star, $M$ is the centrally dominated
mass of the galaxy and $r$ is the radial distance from the center of the galaxy.

However, the actual observations of the rotation curve of spirals completely
disagree with the Eq.(191): the curves do not decrease in the expected inverse
square root relationship. Rather, in most galaxies observed, one finds that
$v$ becomes approximately constant out to the largest values of radial
distance $(r)$ where the rotation curve can be measured (\cite{TOGO18},
P.241). A solution to this problem was to hypothesize the existence of a
substantial invisible amount of matter to account for this inexplicable extra
mass/gravity force that keeps the speed of orbiting stars/gas approximately
constant for extremely large values of $r$. \ This extra mass/gravity was
dubbled "dark matter" \cite{FKA}.

Though dark matter is by far the most accepted explanation of the rotation
problem, other alternatives have been proposed with varying degrees of
success. The most notable of these alternatives is the Modified Newtonian
Dynamics (MOND), which involves modifying the Newton force law by
phenomenologically adding a small fudged factor $\alpha_{0}$:%
\begin{equation}
F_{MOND}=\frac{G_{N}\text{ }Mm}{r^{2}}+\alpha_{0}%
\end{equation}

Within the central bulge of galaxy, the first term of the Eq.(192) dominates,
and to the largest value of $r$ where the rotation curve can be measured (the
domain of dark matter), the second term dominates. MOND has had a remarkable
amount of success in predicting the flat rotation curves of
low-surface-brightness galaxies, matching the Tully-Fisher relation of the
baryonic distribution, and the velocity dispersions of the small orbiting
galaxies of the local group \cite{22}.

The ensuing fundamental question here is: Do we really need to modify
\textbf{Newtonian dynamics} and \textbf{Einstein's GR} before we could account
for this extra gravitational force with no "origin"? The answer is a big No!
The two theories are fantastically accurate in their respective domains of
validity. But the core of the problem is that we assumed that both theories
should be valid at all distance scales (from particle physics scale (say,
Planck scale) to the edge of the Universe); but the irony is that they are
not. It turns out that in order to solve GRP, one needs a force law that is
valid for all distance scales. This is where BCJ construction kicks in. As we
have painstakingly demonstrated (by obtaining Eqs.(106) and (134) from
Eq.(27)) , the major conclusion of double copy construction is the existence
of gauge/gravity duality. This duality property led to the formulation of the
Yang-Mills-Gravity (YM) force in the Eq.(137). A close perusal of Eqs.(137)
and (192) shows that both equations are essentially the same; and that
Eq.(137) explains the \textbf{universal rotation curve perfectly, by producing
a flatly stable curve at large values of }$r$ (see red curve of the
\textbf{Fig.3}) $-$ (a universal rotation curve can be expressed as the sum of
exponential distributions of visible matter that reduce to zero with large
values of $r$ away from the center of galaxy, and spherical dark matter halo
(just like $\sigma$ in the Eq.(137)) that tapers to flat rotation curve with
constant speed and gravitational force \cite{45}) $-$ Hence, the deep
significance of structure of the Eq.(137) to cosmology is not accidental but
fundamental to the evolutionary histories of our universe.

As pointed out by \emph{V. de Sabbata and C. Sivaram}, a deviation from the
Newton's inverse square law can arise naturally from $R+R^{2}$ theory (such as
strong gravity theory), whose solution gives a Newtonian/Coulomb potential and
a Yukawa term (\cite{VDS}, P.4). Thus it is natural to investigate the
behavior of the Eq.(137) on a cosmological scale. Of course, Eq.(137) tapers
to $\sigma$ on the cosmological scale:%

\begin{equation}
F_{YMG}=0.299GeV^{2}=2.449\times10^{5}N
\end{equation}

with mass ($M_{g}$)%
\begin{equation}
M_{g}=\sqrt{F_{YM}}=546.809MeV
\end{equation}

As a result of the Eqs.(137) and (194), the following facts emerge: (1) empty
space/vacuum is permeated with constant-attractive-gravitational force (dark
matter) with mass $M_{g}=546.809MeV$. (2) The dark matter is stable on
cosmological time scales due to the flat curve property of $M_{g}$ for
$r\longrightarrow \infty$ (see red curve in the Fig.3). (3) Newtonian dynamics
and Einstein's GR need no modifications. (4) MOND is phenomenologically
correct and happens to be compatible with YMG force law.

\section{Repulsive Gravity and Cosmic Inflation}

It is true (from the Eq.(137)) that $F_{YMG}$ can only get more repulsive as
we probe shorter and shorter distances. As such, the separating distance
between two gluons cannot taper to zero. This means that the theory "realizes
asymptotic freedom" because two gluons cannot sit on top of each other (i.e.
separating distance $r=0$ is forbidden), hence they are \textbf{almost free}
to move around due to the \textbf{non-existent of attractve force at }%
$r\ll \Lambda_{QCD}^{-1}$. This explanation is then carried by analogy into the
construction of spacetime geometry. The fact that the spacetime metric
($g_{\mu \nu}$) is \emph{ab initio} constructed out of the two entangled gluons
(BCJ construction) means that spacetime \textbf{cannot} realize singularity
(i.e. $r\neq0$). From the foregoing, one is therefore forced to ask a
fundamentally disturbing question: How did our universe "begin", or what
existed "before" the Big Bang?

A. C. Doyle famously claimed that "once you eliminate the impossible, whatever
remain, no matter how improbable, must be the truth." In line with this quote,
we posit that the behavior of the universe during the first fraction of a
second ($t<10^{-44}s$) after the Big Bang can only be a matter for conjecture
but we are certain that $t\neq0$ and $r\neq0$ due to the ever-increasing
repulsive nature of $F_{YMG}$ as we probe short-distance scales. Hence,
perhaps our universe had its origin in the ever-recurring interplay between
\textbf{expanding} and \textbf{contracting} universe. We are sure of the
\textbf{former} but the \textbf{latter }is highly unlikely, given the present
behavior of dark energy and the ever-constant effective Lagrangian of the
vacuum $\upsilon^{2}/M_{X}=3.7meV$.

The fact that most of the calculations done using Planck epoch parameters
(i.e., Planck time, Planck energy and Planck length) conform to what is
obtainable in nature strongly suggests that any epoch less than Planck epoch
is operationally meaningless. Thus, a plausible theory can be constructed
(starting from the Planck time $t=10^{-44}s$) by bringing the calculations
done in the\textbf{\ section IX }of this paper to bear: After about
$10^{-44}s$ \ (with Planck length $10^{-19}GeV^{-1}$) the repulsive gravity
was $F_{YMG}=-2.6\times10^{38}GeV^{2}(=-2.129\times10^{44}N).$ This caused the
universe to undergo an exponential expansion (due to the exponential nature of
$F_{YMG}$). The exponential expansion lasted from $10^{-44}s$ after the Big
Bang/Bounce to time $10^{4}GeV^{-1}(=6.6\times10^{-21}s):$ this is the time
that produced the stable-flat curve (red curve in the Fig.(3)), signaling the
demise of the exponential era. Following this cosmic inflationary (exponential
expansion) epoch, the dark matter $-$ which is nothing but the remnant of
Yang-Mills-Gravity force, $F_{YMG}=2.449\times10^{5}N$ $-$ dominates and the
universe continues to expand but at a less rapid rate. The battle of supremacy
between dark energy (repulsive gravity) and dark matter (attractive gravity)
was won by dark energy during the time $t=6.6\times10^{-21}s$: since dark
energy is intimately connected to the spacetime metric itself (see Eq.(189)),
it would have increased tremendously during the exponential expansion period,
when the space increased in size by factor of $10^{23}(=10^{4}GeV^{-1}%
/10^{-19}GeV^{-1})$ in a small fraction of a second! The victory of dark
energy over dark matter means that our universe will continue to expand
\emph{ad infinitum}: \emph{we are living in a runaway universe.}

Another formal way of explaining the inflationary epoch (i.e. the
vacuum-dominated universe approach) of the universe $-$ which makes the
pre-existing universe scenario conceivable $-$ can be effected by using
\textbf{Fig.3}. In this approach, it is believed that the universe passed
through an early epoch of vacuum dominance (i.e. inflation), presided over by
the varying potential energy (i.e. Eq.(113)) of the scalar field, called
inflaton (\cite{TOGO19}, P.564). When the scalar field reached the minimum of
the potential (which corresponds to the minimum of the potential curve (blue
curve) in the Fig.3) exponential expansion ended. Based on the law of
conservation of energy, the reduction in the potential energy (due to the
rolling down of the inflaton from the top of the potential curve, i.e. the
decaying of the inflaton field) generated hot (quark-gluon) plasma epoch
(which later generated the matter and radiation epoch). So from then on, the
Big Bang evolved according to the Standard Cosmological Model (\cite{TOGO19},
P.564); and governed by the Eqs.(105), (134), (189) and (193).

We conclude this section with the following facts: (1) The initial conditions
of our universe are the Planck epoch parameters (i.e., Planck time, Planck
energy and Planck length). (2) Inflationary theory is correct. (3) The
expansion epoch of the universe consists two phases: (i) the exponential
expansion (governed by $F_{YMG}=-2.129\times10^{44}N$) and the normal
accelerating expansion (governed by Eq.(189)).

\section{Conclusion}

We have shown, in this paper, that the point-like theory of quantum gravity
(strong gravity theory) is geometrically equivalent to the four-dimensional,
nonlinear quantum gauge field theory (i.e. QYMT), and the Einstein General
Relativity. The inherent UV regularity, BCJ and gauge-gravity duality
properties of this renormalizable theory allowed us to solve four of the most
difficult problems in the history of physics: namely, dark matter, existence
of quantum Yang-Mills theory on $R^{4}$, neutrino mass and dark energy problems.

In any geometric field theory, all physical quantities and fields should be
induced from one geometric entity (Weyl's action) and the building blocks of
the geometry used (2-gluon configuration/double-copy construction). This
principle has been inspired by Einstein's statement $-$ "a theory in which the
gravitational field and electromagnetic field do not enter as logically
distinct structures, would be much preferable"$-$ and established in this
paper. As we have demonstrated, this principle implies that the Weyl
Lagrangian density used to construct the field equations of the strong gravity
theory is composed of the building blocks (two gluons) of the geometry and
their derivatives (in which the curvature arised in terms of derivatives of
the dressed gluon field). In other words, Weyl Lagrangian is not constructed,
\emph{a priori}, \textbf{from different parts} (each corresponding to a
certain field) as usually done. This makes strong gravity theory to pass the
test of unification principle.

\textbf{Acknowledgement}

Mr. O. \  \ F. Akinto is indebted to the Department of Physics, CIIT, Islamabad
\ and the National Mathematical Center Abuja, Nigeria \ for their financial support.

\bigskip

\bigskip


\begin{thebibliography}{999}                                                                                              %
\bibitem {CJI}C. J. Isham, A. Salam and J. Strathdee, "2$^{+}$ Nonet as Gauge
particles for $S%
\mathcal{L}%
(6,C)$ symmetry", Phys. Rev. \textbf{D8,} 8 (1973).

\bibitem {ASJ}A. Salam and J. Strathdee, "Class of solutions for the
strong-gravity equations", Phys. Rev. \textbf{D16, }8 (1977).

\bibitem {CSI}C. Sivaram and K. P. Sinha, "Strong Spin-two Interaction and
General Relativity", Phys. Rep. (Rev. Sect. Phys. Lett.) \textbf{51, }3 (1979).

\bibitem {DJS}Dj. Sijacki and Y. Ne'eman, "QCD as an effective strong
gravity", Phys. Lett. \textbf{B247, }4 (1990).

\bibitem {YNE}Y. Ne'eman and Dj. Sijacki, "Proof of pseudo-gravity as QCD
approximation for the hadron IR region and $J\sim M^{2}$ Regge trajectories",
Phys. Lett. \textbf{B276} (1992).

\bibitem {ASCS}A. Salam and C. Sivaram, " STRONG GRAVITY APPROACH TO QCD AND
CONFINEMENT", Mod. Phys. Lett. \textbf{A8}, 4 (1993).

\bibitem {IANC}I. Antoniadis and N. C. Tsamis, "Weyl Invariance and the
Cosmological Constant", SLAC-PUB-3297, (1984).

\bibitem {KSS}K. S. Stelle, "Renormalization of higher-derivative quantum
gravity", Phys. Rev. \textbf{D16, }4 (1977).

\bibitem {EW}E. Witten, " Physical law and the quest for mathematical
understanding", Bulletin (New Series) of the American Mathematical Society,
\textbf{40, }1\textbf{\ }(2002).

\bibitem {JCA}J. Carson, A. Jaffe and A. Wiles (Editors),"The Millennium Prize
Problems", American Mathematical Society (2006).

\bibitem {BIF}B. I. Loffe, V. S. Fadin and L. N. Lipatov, "Quantum
Chromodynamics: Perturbative and Non-perturbative Aspects", Cambridge
University Press, New York (2010).

\bibitem {QHN}J. Goity, C. Keppel and G. Prezeau (Editors),"Proceedings of the
16th and 17th Annual Hampton University Graduate Studies", World Scientific
Publishing Co. Pte. Ltd (2001 \& 2002).

\bibitem {JBER}J. Beringer et al., (Particle Data Group), Phys. Rev.
\textbf{D86, }010001 (2012).

\bibitem {TOGO1}H. Kawai, D. C. Lewellen and S. H. H. Tye, "A relation between
tree amplitudes of closed and open strings", Nucl. Phys. \textbf{B269}, 1 (1986).

\bibitem {TOGO2}Z. Bern, J. J. M. Carrasco and H. Johansson, "New Relations
for Gauge-Theory Amplitudes", Phys. Rev. \textbf{D78}, 085011 (2008).

\bibitem {TOGO3}Z. Bern, J. J. M. Carrasco and H. Johansson, "Perturbative
Quantum Gravity as a Double Copy of Gauge Theory", Phys. Rev. Lett.
\textbf{105}, 061602 (2010).

\bibitem {TOGO4}American Physical Society, "J. J. Sakurai Prize for
Theoretical Particle Physics", 2014 (Official site).

\bibitem {TOGO5}M. Chiodaroli, M. Gunaydin, H. Johansson and R. Roiban,
"Scattering amplitudes in N = 2 Maxwell-Einstein and Yang-Mills-Einstein
supergravity", JHEP \textbf{1501}, 081 (2015); [arXiv:1512.09130 [hep-th]];
[arXiv: 1511.01740 [hep-th]].

\bibitem {TOGO6}H. Johansson and R. Roiban, "Simplifying Multiloop Integrands
and Ultraviolet Divergences of Gauge Theory and Gravity Amplitudes", Phys.
Rev. \textbf{D85}, 105014 (2012).

\bibitem {TOGO7}D. Hilbert, "MATHEMATICAL PROBLEMS", Bulletin (New Series) of
the American Mathematical Society \textbf{37}, 4 (2000).

\bibitem {TOGO8}F. Bissey, et al., "Gluon flux-tube distribution and linear
confinement in baryons", Phys. Rev. \textbf{D76}, 114512 (2007); D. B.
Leinweber, "Visualizations of Quantum Chromodynamics", Centre for the
Subatomic Structure of Matter (CSSM) and Department of Physics, University of
Adelaide, 5005 Australia (2003, 2004).

\bibitem {TT}J. T. Ttrujillo, "Weyl Gravity as a Gauge Theory": All Graduate
Theses and Dissertations, Paper 1951, Utah State University (2013).

\bibitem {SW-2}S. Weinberg, "Gravitation and Cosmology: Principles and
Applications of the General Theory of Relativity", Massachusetts Institute of
Technology, John Wiley and Sons (1972).

\bibitem {TOGO9}B. S. DeWitt, "In Relativity Groups and Topology", ed. C.
DeWitt and B. DeWitt (Gordon and Breach, 1964), P.719.

\bibitem {CWKJ}C. W. Misner, K. S. Thorne and J. A. Wheeler, "Gravitation", W.
H. Freeman and Company, New York (1973).

\bibitem {VDS}Venzo de Sabbata and C. Sivaram, "Fifth Force as a Manifestation
of Torsion", Int. J. Theor. Phys., \textbf{29}, 1 (1990).

\bibitem {TOGO10}H. D. Politzer, "Reliable Perturbative Results for Strong
Interactions", Phys. Rev. Lett. \textbf{30 }(1973), 1346-1349; Phys. Rep.
\textbf{14C}, 129 (1974).

\bibitem {TOGO11}D. J. Gross and F. Wilczek, "Ultraviolet Behavior of
Non-Abelian Gauge Theories", Phys. Rev. Lett. \textbf{30} (1973), 1343-1346;
Phys. Rev. \textbf{D8}, 3633 (1973); \textbf{D9}, 980 (1974).

\bibitem {TOGO12}D. Gross, "Methods in Field Theory", ed. R. Balian and J.
Zinn-Justin, North-Holland (1976).

\bibitem {TOGO13}W. Marciano and H. Pagels, "Quantum chromodynamics", Phys.
Rep. \textbf{36C}, 3 (1978).

\bibitem {TOGO14}J. W. Rohif, "Modern Physics from a to z0", Wiley, (1994).

\bibitem {TOGO15}D. Ashok and F. Thomas, "Introduction to Nuclear and Particle
Physics", Wiley, (1994).

\bibitem {PCR}P. C. Riedi, "Thermal Physics", Macmillan Press Ltd, 1st ed. (1976).

\bibitem {CSB}O. Kiriyama, M. Maruyama and F. Takagi, "Current quarks mass
effects on chiral phase transition of QCD in the improved ladder
approximation", Phys. Rev. \textbf{D62, }105008 (2000).

\bibitem {SER}S. E. Rugh and H. Zinkernagel, "The quantum vacuum and
cosmological constant problem", Studies in History and Philosophy of Science
Part \textbf{B33}, 4 (2002).

\bibitem {LZW}L. Zhou, M. Weixing and L. S. Kisslinger, "Theoretical
prediction of cosmological constant $\Lambda$ in Veneziano ghost theory of
QCD", J. Mod. Phys. \textbf{3} (2012).

\bibitem {FRU}F. R. Urban and A. R. Zhitnitsky, "The cosmological constant
from the QCD Veneziano ghost", Phys. Lett. \textbf{B688}, 9 (2010).

\bibitem {SWEN}S. Weinberg, "The cosmological constant problem", Rev. Mod.
Phys. \textbf{61, }1 (1998).

\bibitem {AHM}A. H. Mueller, Phys. Rep. 73C (1981) 237; A. J. Buras, Rev. Mod.
Phys. 52 (1980) 199; S. J. Brodsky, T. Huang, G. P. Lepage, SLAC-PUB-2868,
"Quarks and Nuclear Forces", Springer, Vol. (100), (1982).

\bibitem {GPL}G. P. Lepage and S. J. Brodsky, "Exclusive processes in
perturbative quantum chromodynamics", Phys. Rev. \textbf{D22}, 2157 (1980).

\bibitem {ASO}A. Jaffe, G. Parisi and D. Ruelle (Editors)," Workshop on
Non-perturbative Quantum Chromodynamics", Birkhauser Boston, Inc. (1983).

\bibitem {GDISS}S. Bethke, G. Dissertori and G. P. Salam, " Quantum
Chromodynamics", Princeton University Press (2012).

\bibitem {VGKA}V. G. Krivokhijine, A. V. Kotikov, "A systematic study of QCD
coupling constant from deep inelastic measurements", Physics of Atomic Nuclei,
\textbf{68}, 11 (2005); [arXiv:0108224 [hep-ph]] (2001).

\bibitem {TOGO16}A. Pich, "Effective Field Theory", Elsevier Science B. V.
(2008), P.11.

\bibitem {TOGO17}H. Georgi and M. Machacek, "Doubly charged Higgs bosons",
Nucl. Phys. \textbf{B262}, 3 (1985); M. A.S. Kanemura, M. Kikuchi and K.
Yagyu, Phys. Lett. \textbf{B714, }279 (2012).

\bibitem {TOGO18}K. A. Olive et al. (Particle Data Group), "Particle Physics
Booklet", Chin. Phys. \textbf{C38}, 09001 (2014) P.221.

\bibitem {TOGO19}C. Giunti and C. W. Kim, "Fundamental of Neutrino Physics and
Astrophysics", Oxford University Press (2007) P.207.

\bibitem {TOGO20}R. A. Diaz, D. Gallego and R. Martinez, Int. J. Mod. Phys.
\textbf{A22 }(2007), 1849-1874.

\bibitem {TOGO21}Nikolai Kochelev and Dong-Pil Min, "Role of glueballs in
non-perturbative quark-gluon plasma", Phys. Lett. \textbf{B650} (2007), 239-243.

\bibitem {AAKS}A. A. Khan, S. Aoki, et al., "Light hadron spectroscopy with
two flavors of dynamical quarks on the lattice", Phys. Rev. \textbf{D65},
054505 (2002).

\bibitem {DSE}Shu-Sheng Xu, et al., "The chiral phase transition with a
chemical potential in the framework of Dyson-Schwinger equations", Phys. Rev.
\textbf{D91}, 056003 (2015).

\bibitem {UAG}U. Aglietti, Phys. Lett. B281, 341 (1992); L. Trentadue, Act.
Phys. Pol. B38 (2007) 11.

\bibitem {MGA}V. Gogokhia \ and G. G. Barnafoldi, " The Mass Gap and its
Applications", \ World Scientific Publishing Co. Pte. Ltd (2012).

\bibitem {DGR}D. Griffiths, "Introduction to Elementary Particles", John Wiley
\& Sons (1987).

\bibitem {CITZ}C. Itzykson and J. B. Zuber, "Quantum Field Theory",
McGraw-Hill (1980).

\bibitem {SW-3}L. A. Gaume and M. A. V. Mozo, "An Invitation to Quantum Field
Theory", Springer (2012).

\bibitem {SW-4}P. J. Mohr, B. N. Taylor and A. B. Newell, "CODATA recommended
values of the fundamental physical constants", Rev. Mod. Phys. \textbf{84},
1527 (2012).

\bibitem {SW-5}R. Horsley, et al., Phys. Lett. \textbf{B732} (2014).

\bibitem {JEFF}J. Greensite, " Lecture Notes in Physics 821: An Introduction
to the Confinement Problem", Springer (2011).

\bibitem {ANIVA}A. N. Ivanov, N. I. Troitskaya, M. Faber and M. Nagy, "Chiral
symmetry breaking in QCD with linearly rising confinement potential", II Nuovo
Cimento \textbf{A107} (9), (1994).

\bibitem {TOGO22}M. Thomson, "Modern Particle Physics", Cambridge University
Press, (2013).

\bibitem {TOGO23}A. Deur, S. J. Brodsky and G. F. de Teramond, "The QCD
Running Coupling", SLAC-PUB-16448 (JLAB-PHY-16-2199) 2016, P.9;
[arXiv:1604.08082 [hep-ph]].

\bibitem {TOGO24}D. V. Bugg (ed.)," Hadron Spectroscopy and the Confinement
Problem", NATO ASI Series B: Physics Vol. \textbf{353, }Plenum Press New York
and in NATO Scientific Affairs Division (1995).

\bibitem {TOGO25}S. M. Carrol, "Spacetime and Geometry", Addison-Wesley (2004).

\bibitem {TOGO26}A. V. Manohar and H. Georgi, "Chiral quarks and the
nonperturbative quark model", Nucl. Phys. \textbf{B234}, 189 (1984).

\bibitem {MCORN}J. M. Cornwall, "What is the relativistic generalization of a
linearly rising potential?", Nucl. Phys. \textbf{B127}, 75(1977).

\bibitem {CGATT}C. Gattringer and C. B. Lang, "Qauntum Chromodynamics on the
Lattice: An Introductory Presentation", Springer (2010).

\bibitem {HJW}H. J. W. Muller and K. Schilcher, "High-energy scattering for
Yukawa potentials", J. Math. Phys.\textbf{\ 9} (1968); H. Yukawa, "On the
interaction of elementary particles", Proc. Phys. Math. Soc. \textbf{17, }48
(1935); G. E. Brown and A. D. Jackson, "The Nucleon-Nucleon Interaction",
North-Halland Publishing, Amsterdam (1976).

\bibitem {DAVID}D. Scherfgen, htt:// www.derivative-calculator.net.

\bibitem {HENG}H. T. Ding,"Recent Lattice QCD results and phase diagram of
strongly interacting matter", Nucl. Phys. \textbf{A931} (2014).

\bibitem {TOGO27}A. Jaffe and E. Witten, "Quantum Yang-Mills Theory: Official
problem description", Clay Mathematics Institute (2000).

\bibitem {TOGO27A}L. O'Raifeartaigh, "The Dawning of Gauge Theory", Princeton
University Press, (1997); C. N. Yang and R. L. Mills, "Conservation of
isotopic spin and isotopic gauge invariance", Phys. Rev. \textbf{96 }(1954),
191-195\textbf{.}

\bibitem {TOGO28}T. Rodrigo and A. Ruiz (eds.), "Proceedings of the XXIII
International Meeting on Fundamental Physics", World Scientific (1995).

\bibitem {EFEMI3}E. Bertuzzo and C. Frugiuele, "Natural SM-like 126GeV Higgs
boson via non-decoupling D terms", Phys. Rev. \textbf{D93, }035019 (2016).

\bibitem {EFEMI4}Bin He, N. Okada and Qaisar Shafi, "125GeV Higgs, type III
seesaw and gauge-Higgs unification", Phys. Lett. \textbf{B716 }(2012).

\bibitem {FRAG}F. Englert, "Nobel Lecture: The BEH mechanism and its scalar
boson", Rev. Mod. Phys. \textbf{86, }843 (2014).

\bibitem {IATO}I. Antoniadis and D. Ghilencea (Editors), "Supersymmetry After
the Higgs Discovery", Springer (2014).

\bibitem {FBP}F. Boehm and P. Vogel, "Physics of Massive Neutrinos", 2nd Ed.,
University of Cambridge Press (1987).

\bibitem {SWJ}S. Wojcickl, " Prospects in Neutrino Physics", University of
Stanford Press (1996).

\bibitem {TKAB}T. Kajita and A. B. McDonald, "The Nobel Prize in Physics
2015", Nobel prize.org (2015).

\bibitem {TKAA7}A. Ereditato, T. Kajita and A. Masiero, New J. Phys. \textbf{9
}(2007); T. Kajita, New J. Phys. \textbf{6, }194 (2004); J. Phys. G (Nucl.
Part. Phys. \textbf{29}, 1471 (2003)); J. Phys.(Conf. Ser. \textbf{308,
}012003 (2011)); J. Phys. (Conf. Ser. \textbf{136}, 022020 (2008)); T. Kajita,
M. Ishitsuka, H. Minakata and H. Nunokawa, J. Phys.: Conf. Ser. \textbf{39},
332 (2006).

\bibitem {ABMD3}A. B. McDonald, New J. Phys. 6 (2004) 121; Phys. Scr. 2006, 10
(2006); J. Phys.(Conf. Ser. 173 (2009) 012002).

\bibitem {ASNE}S. N. Ahmed, et al., (SNO Collaboration), Phys. Rev. Lett.
(2003) 041801.

\bibitem {KEG}K. Eguchi, et al., (KamLAND Collaboration), Phys. Rev. Lett. 90
(2003) 021802.

\bibitem {PAD}P. Adamson, et al., (MINOS Collaboration), Phys. Rev. Lett. 107
(2011) 181802.

\bibitem {YABD}Y. Abe, et al., (Double Chooz Collaboration), Phys. Rev. Lett.
108 (2012) 131801; Phys. Rev. D86 (2012) 052008.

\bibitem {FANE}F. An. et al., (Daya Bay Collaboration), Phys. Rev. Lett. 108
(2012) 171803; F. An, et al., (Daya Bay Collaboration), Chin. Phys. C37 (2013) 011001.

\bibitem {JAHN}J. Ahn, et al., (RENO Collaboration), Phys. Rev. Lett. 108
(2012) 191802.

\bibitem {YVKO}Y. V. Kozlov, V. P. Martem 'yanov and K. N. Mukhin, Phys. 40
(1997) 8.

\bibitem {DVACB}D. V. Ahluwalia and C. Burgard, Phys. Rev. D57 (1998) 8; Gen.
Rel. Grav. 28 (1996).

\bibitem {DBAL}D. Balin and A. Love, "Supersymmetric Gauge Field Theory and
String Theory", CRC Press (1994).

\bibitem {MIW}M. I. Wanas, Int. J. Theor. Phys. 24 (1985); Stud. Cercet. Stin.
Ser. Mat. 10 (2001); Int. J. Geom. Meth. Mod. Phys. Lett. A25 (2010).

\bibitem {FIMMW}F. I. Mikkhial, M. I. Wanas and G. G. L. Nashed, Astrophys. \&
Space Sci, 228 (1995): Proceedings of the fourth United Nations / European
Space Agency Workshop, Cairo.

\bibitem {GEOLE}G. Lemaitre, "Quaternions et e t espace elliptique", Act.
Pont. Acad. Sc., 12 (1948).

\bibitem {MAAPE}M. K. Peterson, "Dante and the 3-Sphere", Americ. J. Phys., 47
(1979) 12.

\bibitem {MKMR}M. Kadastik, M. Raidal and L. Rebane, Phys. Rev. D77 (2008) 115023.

\bibitem {HNWJ}H. Nunokawa, W. J. C. Teves and R. Z. Funchal, Phys. Lett. B
562 (2003) 1.

\bibitem {HVKLK}V. Klapdor-Kleingrothaus, " Seventy years of Double Beta Decay
from Nuclear Physics to Beyond Standard Model Particle Physics", World
Scientific Publishing Co. Pte Ltd (2010); R. Mohaptra, A. Smirnov, Ann. Rev.
Nucl. Part. Sci. 56 (2006) 569; M. Czakon, J. Gluza, J. Studnik and M. Zralek,
Phys. Rev. D65 (2001) 5; P. Huber, " Neutrino Theory", Center for Neutrino
Physics at Virginia Tech., International Workshop on Baryon and Lepton Number
Violation, University of Massachusetts (2015).

\bibitem {JBDH}J. Bielefeld, D. Huterer and E. V. Linder, JCAP05 (2015) 023;
P. A. R. Ade, et al., Astron. \& Astrophys. manuscript no. draft P1011 (2013)
\& manuscript no Planck Lensing (2015).

\bibitem {SBDCQ}S. Bocquet, et al., APJ 799 (2015) 2.

\bibitem {AGRI}B. P. L. Ward, "CURRENT STATUS OF LHC PHYSICS: PRECISION THEORY
VIEW", ACTA PHYS POL \textbf{B45}, 7 (2014) P. 1632; A. G. Riess et al.,
Astron. J. \textbf{116}, 1009(1998); S. Perlmutter et al., Astrophys.
J.\textbf{\ 517}, 565 (1999); and references therein.

\bibitem {AGRI1}M. J. Luo, "The cosmological constant problem and
re-interpretation of time", Nucl. Phys. \textbf{B884} (2014).

\bibitem {V.R}V. Rubin and W. K. Ford, Jr, "Rotation of the Andromeda Nebula
from a Spectroscopic Survey of Emission Regions", Astrophys. J. \textbf{159},
379 (1970); V. Rubin, N. Thonnard and W. K. Ford Jr, "Rotational Properties of
21 SC Galaxies with a Large Range of Luminosities and Radii from NGC 4605 (R =
4kpc) to UGC 2885 (R = 122kpc)", Astrophys. J.\textbf{\ 238}, 471 (1980).

\bibitem {ECO}E. Corbelli and P. Salucci, "The extended rotation curve and the
dark matter halo of M33", Monthly Notices of the Royal Astron. Soc.
\textbf{311}, 2 (2000).

\bibitem {VTR}V. Trimble, "Existence and nature of dark matter in the
universe", Annual Review of Astron. and Astrophys. \textbf{25} (1987).

\bibitem {FKA}F. Katherine, "The Cosmic Cocktail: Three Parts Dark Matter",
Princeton University Press (2014).

\bibitem {22}S. S. McGaugh and W. J. G. de Block, "Testing the Hypothesis of
Modified Dynamics with Low Surface Brightness Galaxies and other Evidence",
Astrophys. J. \textbf{499}, 1 (1998); S. S. McGaugh, "Novel Test of Modified
Newtonian Dynamics with Gas Rich Galaxies", Phys. Rev. Lett. \textbf{106}, 12
(2011); S. S. McGaugh and M. Milgrom, "Andromeda Dwarfs in Light of Modified
Newtonian Dynamics", Astrophys. J. \textbf{766}, 1 (2013).

\bibitem {45}P. Salucci and A. Borriello, "The Intriguing Distribution of Dark
Matter in Galaxies", Lecture Notes in Physics 616: [arXiv:020345 [atroph-ph]].
\end{thebibliography}
\end{document}